\documentclass[12pt,a4paper]{iopart}
\usepackage{amssymb}
\usepackage{epsfig}
\usepackage{multirow}
\usepackage{graphicx}
\usepackage{wasysym}

\newcommand {\dr}{{\mathrm d}\mathbf{r}}

\newcommand {\dx}{{\mathrm d}\mathbf{x}}
\newcommand {\dd}{{\mathrm d}}
\newcommand {\rr}{\mathbf{r}}
\newcommand {\drr}{{\mathrm d}\mathbf{r}}

\newcommand {\xx}{\mathbf{x}}
\newcommand{\suprm}[1]{\ensuremath{^{\mathrm{#1}}}}
\newcommand{\im}{\ensuremath{{\mathrm{i}}}}
\newcommand{\subrm}[1]{\ensuremath{_{\mathrm{#1}}}}
\newcommand{\sgn}{\mathop{\mathrm{sgn}}}
\newcommand{\tfrac}[2]{\frac{#1}{#2}}

\begin{document}

\title[Non-additive hard sphere mixtures]{Binary non-additive hard sphere mixtures: Fluid demixing, asymptotic decay of correlations and free fluid interfaces}

\author{Paul\ Hopkins}
\address{H.H.\ Wills Physics Laboratory, University of Bristol, Tyndall Avenue, Bristol BS8 1TL, UK}
\ead{Paul.Hopkins@bristol.ac.uk}

\author{Matthias\ Schmidt}
\address{Theoretische Physik II, Physikalisches Institut, Universit{\"a}t Bayreuth, D-95440 Bayreuth, Germany}
\address{H.H.\ Wills Physics Laboratory, University of Bristol, Tyndall Avenue, Bristol BS8 1TL, UK}

\date{\today}

\begin{abstract}
Using a fundamental measure density functional theory we investigate both bulk and inhomogeneous systems of the binary non-additive hard sphere model. For sufficiently large (positive) non-additivity the mixture phase separates into two fluid phases with different compositions. We calculate bulk fluid-fluid coexistence curves for a range of size ratios and non-additivity parameters and find that they compare well to simulation results from the literature. Using the Ornstein-Zernike equation, we investigate the asymptotic, $r\to\infty$, decay of the partial pair correlation functions, $g_{ij}(r)$. At low densities there occurs a structural crossover in the asymptotic decay between two different damped oscillatory modes with different wavelengths corresponding to the two intra-species hard core diameters. On approaching the fluid-fluid critical point there is Fisher-Widom crossover from exponentially damped oscillatory to monotonic asymptotic decay. Using the density functional we calculate the density profiles for the planar free fluid-fluid interface between coexisting fluid phases. We show that the type of asymptotic decay of $g_{ij}(r)$ not only determines the asymptotic decay of the interface profiles, but is also relevant for intermediate and even short-ranged behaviour. We also determine the surface tension of the free fluid interface, finding that it increases with non-addivity, and that on approaching the critical point mean-field scaling holds.
\end{abstract}


\maketitle

\section{Introduction}
\label{sec:intro}
Liquids can consist of a mixture of components which in molecular systems may be different atomic or molecular components. In colloidal systems mixtures may be formed from particles with differing shapes or sizes, or mixtures of colloids and non-adsorbing polymer. Besides gas-liquid phase separation such mixtures may exhibit liquid-liquid separation, where the system demixes into two (or more) phases with differing compositions. 

Inter-particle interaction potentials between the constituent particles in a gas or a liquid may contain both short-range repulsion, due to the overlap of outer electron shells, and longer-ranged attractive or repulsive tails, due to dispersion or Coloumb forces~\cite{hansen2006theory}. For intermediate densities both features of the potential are important, and  as van der Waals discovered~\cite{van2004continuity}, it is the presence of an attractive tail that drives liquid-gas phase separation. However, if the liquid is dense, the long-range tail becomes less important and the structure of the fluid is primarily determined by the short-range repulsion.

Following van der Waals~\cite{van2004continuity} it is convenient to separate the potential into its short-range sharply repulsive and longer-range components and treat them within a theoretical approach separately. The simplest model for the short-ranged repulsive potentials is the hard sphere model which disallows particle overlap. This model has been shown to give a good approximation to the thermodynamic and structural properties of fluids, particularly near crystallisation. The development of ever better approximate theoretical treatments of the hard sphere model is a major element of liquid-state theories. Furthermore, given a theoretical treatment of the hard sphere potential, attractive or repulsive tails may be incorporated using relatively simple perturbation theories. %The structure of dense fluids is largely determined by short-range repulsive forces between particles. The simplest model of these potentials is the hard-core model which disallows particle overlap. This model has been shown to give a good approximation to the thermodynamic and structural properties of fluids, particularly near crystallisation where the structure of the fluid does not differ from that with more complicated potentials~\cite{hansen2006theory}. Furthermore, given a theoretical hard sphere mode attractive or repulsive tails may be incorporated into the theory using relatively simple perturbation theories.
It is straightforward to generalise the hard sphere model to multi-component mixtures. This simplest mixture has two components where the second species can have the same or a different diameter than the first species. This two-component mixture is normally formulated such that the distance of closest approach between particles of different species is a simple mean of the diameters of the particles of each species. In a real fluid this assumption may not be true and relaxing this constraint allows novel features of real systems to be investigated~\cite{louis2000cap,roth2001theory,santos2005eos}.

Specifically, we define a two-component mixture of particles that interact through the hard sphere pair-potentials,
\begin{eqnarray}
\label{cases}
V_{ij}(r)=\cases{
\infty & $r<\sigma_{ij}$\\
0 & $\mathrm{otherwise}$,}
\end{eqnarray}
where $i,j=1,2$ label the species, and $\sigma_{ii}$ are the particle diameters. The usual additive cross-species range of interaction is $\sigma_{12}=\tfrac{1}{2}(\sigma_{11}+\sigma_{22})$ but the non-additive hard sphere (NAHS) model generalises this so that $\sigma_{12}$ can be smaller or larger than the arithmetic mean of the like-species diameters,
\begin{equation}
\sigma_{12}=\tfrac{1}{2}(1+\Delta)(\sigma_{11}+\sigma_{22}),
\label{eq:sigma12}
\end{equation} 
where $\Delta\geq-1$ measures the degree of non-additivity. We characterise the model by the size ratio, $q=\sigma_{11}/\sigma_{22}\leq1$, and by $\Delta$. For $\Delta=0$ the model reverts to the binary additive hard sphere model.

If both $\Delta$ and the density of the fluid are sufficiently large, the fluid demixes into two phases, one rich in particles of species 1, one rich in particles of species 2~\cite{louis2000cap,frenkel1994sep,dijkstra1998pbn,gozdz2003cpa,jagannathan2003mcs}.  Experimental work on a number of systems, including alloys, aqueous electrolyte solutions, and molten salts, suggests that non-additivity may lead to both hetero-coordination and homo-coordination~\cite{santos2005eos,ballone1986additive}. In recent work Kalcher \etal have used Monte Carlo simulations to calculate an effective interaction between charged ions in a electrolyte solution~\cite{PhysRevLett.104.097802}, integrating out the degrees of freedom of the solvent molecules. Using a Barker-Henderson mapping they identified hard core diameters with values of non-additivity as large as 0.36 for NaI. In contrast, for $\Delta<0$ mixing of the two species is encouraged and the fluid can exhibit strong short-range order~\cite{gazzillo1989chemical}. For $\Delta$ being sufficiently negative, clustering effects, mesoscopic ordering and the formation of heterogenous structures were reported~\cite{gazzillo1989chemical,gazzillo1990role,hoffman2006mst}.

Besides the additive model there are two further important limiting cases of the binary NAHS model. The Asaka-Oosawa-Vrij model~\cite{asakura1954oib,vrij1976pai,dijkstra1999phase} has long been used as a simple description for the behaviour of a mixture of colloids and non-adsorbing polymers. The colloids interact through hard sphere interaction with diameter $\sigma_{cc}$, while the polymers are treated as ideal, $\sigma_{pp}=0$. The colloid-polymer interaction is also hard sphere like but with a diameter $\sigma_{cp}=\sigma_{cc}/2+R_g$, where $R_g>0$ is the polymer radius of gyration. In the formulation of~\eref{eq:sigma12} this corresponds to $\Delta=2R_g/\sigma_{cc}$. The second important case is the binary Widom-Rowlinson model~\cite{widom1970nmf} where both same-species interactions are ideal, $\sigma_{11}=\sigma_{22}=0$, but the cross-species interaction is through a hard-core with diameter $\sigma_{12}>0$. This corresponds to taking the limit of \eref{eq:sigma12} where $\Delta\to\infty$ and $\sigma_{11}=\sigma_{22}\to0$ while keeping $\sigma_{12}$ constant. The Widom-Rowlinson model has become an important model in statistical physics due to its entropy driven demixing transition, yet simple structure of interactions. For a more thorough review of the literature on NAHS mixtures we refer the reader to the review article~\cite{santos2005eos} and to the more recent contributions~\cite{PhysRevE.74.021106,pellicane2007virial,santos:204506,sillren2010critical}.

% This binary NAHS model has been used as a convenient approximation for a number of condensed matter systems, including colloid-colloid and colloid-star polymer mixtures. 

% One limit of the model, the Asaka-Oosawa-Vrij model~\cite{asakura1954oib,vrij1976pai}, has long been used as a good description for the behaviour of a mixture of colloids and polymers. The colloids interact through a hard sphere interaction with diameter $\sigma_{cc}$, while the polymers are treated as ideal, $\sigma_{pp}=0$, the colloid-polymer interaction is also hard sphere but with a diameter $\sigma_{cp}=\sigma_{cc}/2+R_g$, where $R_g>0$ is the polymer's radius of gyration. For sufficiently large densities the mixture separates into two fluid phases, one rich in colloids, and the other rich in polymers. The degrees of freedom of the polymers can also be integrated out to produce an effective interaction between colloids which now features an attractive well with a depth and range dependent on the polymer density and $R_g$. Another limit of the non-additive hard sphere model is the binary Widom-Rowlinson model~\cite{widom1970nmf} where both same-species interactions are ideal, $\sigma_{11}=\sigma_{22}=0$, but the cross-species interaction is through a hard-core with diameter $\sigma_{12}>0$.

In this paper we explore the binary NAHS using the recently developed fundamental measure density functional theory for this model~\cite{schmidt2004rfn}. Fundamental measure theories (FMTs) are a class of density functional theories (DFTs) that are based on the geometrical quantities (fundamental measures) of the particles involved, e.g.~volume, surface area, radius, and Euler characteristic. These fundamental measures enter the theory through weight functions which are based on these measures. The weight functions are convolved with the density profiles in order to give a set of weighted densities which are then combined within a free energy density. The original functional was formulated by Rosenfeld for additive mixtures of hard spheres~\cite{rosenfeld1989free} using both scalar and vectorial weight functions. Subsequently Kierlik and Rosinberg constructed a functional for hard spheres using only scalar weight functions~\cite{kierlik1990free}. Their theory was later shown to be equivalent to Rosenfeld's formulation~\cite{phan1993equilvalence}. These functionals reproduce the Percus-Yevick approximation for the two-body correlation functions in the bulk fluid. One drawback of these original hard sphere FMTs was that they were not suitable for studying crystallisation phenomena due to unphysical divergences within the functional when the density profiles became strongly confined. This was later remedied (for hard spheres), first using a simple modification~\cite{rosenfeld1997fundamental}, and then later by introducing tensorial weight functions~\cite{tarazona2000density}. For recent reviews on DFT and hard body DFTs in particular, we refer the reader to~\cite{tarazona2008density,roth2010fundamental}. In studies that preceded the NAHS functional, FMTs  were proposed for both the Asakura-Oosawa~\cite{schmidt2000colloidpolymer} and Widom-Rowlinson~\cite{schmidt2000widomrowlinson} models. There is a number of publications dedicated to the study of interfacial properties of the Asakura-Oosawa model~\cite{brader2000fluid,brader2003statistical,wessels2004wall}. Note that the present DFT reduces to that used in~\cite{brader2003statistical,wessels2004wall}, when the Asakura-Oosawa limit of the general non-additive hard sphere mixture is taken.

The NAHS functional, which was first introduced in~\cite{schmidt2004rfn} is based on the  scalar Kierlik-Rosinberg deconvolution~\cite{kierlik1990free} of the hard sphere weight functions, but introduces a further ten scalar weight functions to take account of the non-additivity. It was shown that the functional correctly predicts fluid-fluid phase separation and that the theory provides a reasonable prediction for the location of the critical point compared to existing simulation results. Using the Ornstein-Zernike (OZ) equation, i.e. by inverse Fourier transforming the analytic (Fourier space) total correlation functions, it was shown that the theory provides good account of the radial distribution functions, $g_{ij}(r)$, as compared to Monte Carlo simulation results, though these are not the same as those obtained by the Percus-Yevick (PY) approximation and they violate the core condition $g_{ij}(r)=0$ for $r<\sigma_{ij}$, where $g_{ij}(r)$ is the partial pair correlation function between species $i$ and $j$. The non-additive functional has also been formulated for the one-dimensional version of the model, binary non-additive rods on a line~\cite{schmidt2007fmd}, making accurate predictions for the particle correlation functions, although failing to reproduce the exact solution~\cite{santos2007ebc}. In further work~\cite{schmidt2007poc}, the spherical and one-dimensional convolution transforms in the theory were investigated and shown to form an Abelian group.

In more recent work, Ayadim and Amokrane~\cite{ayadim2010generalization} have used the functional of~\cite{schmidt2004rfn} to calculate the radial distribution functions via the Percus test particle route~\cite{percus1962approximation}. This involves introducing an external potential that represents a single particle fixed at the origin and numerically solving for the density profiles around it. The core conditions are automatically satisfied. These results for the radial distribution functions can then be compared to those obtained from the simpler OZ route, in order to assess the internal consistency of the functional. The authors of~\cite{ayadim2010generalization} found that $g_{ij}(r)$ calculated via the test particle route exhibit small but clearly noticeable unphysical jumps that are not present in the results from the OZ route.They argue that these are not numerical artifacts, but are due to shortcomings in the construction of the functional, and suggest that the functional requires changes at a fundamental level to eliminate the occurrence of discontinuities. In the present investigation in planar (rather than spherical) geometry, we do not find unphysical kinks in density profiles. Both planar fluid-fluid interfaces, as well as the density profiles near a planar hard wall~\cite{hopkins2010slit} are free of artifacts. Furthermore, the current study is dedicated to the intermediate and long-ranged behaviour of the bulk fluid pair correlation functions. We expect this to be largely unaffected by the jumps, which were found to occur at short separation distances~\cite{ayadim2010generalization}.

For fluids where the pair-potentials are short-ranged, it can be shown that $h_{ij}(r) = g_{ij}(r)-1$ can be evaluated by determining the positions and residues of the poles (divergences) of the complex structure factors, $S_{ij}(k)$, where $k$ is a complex wave-number~\cite{evans1994adc}. These poles either occur as a complex conjugate pair, which give rise to a damped oscillatory contribution to $rh_{ij}(r)$, or as a single purely imaginary pole, which gives rise to a purely exponentially decaying contribution. In general, there is an infinite number of poles. However, in order to find the intermediate and asymptotic, $r\to\infty$, decay of $h_{ij}(r)$, it is usually sufficient to find the positions of a small number of poles -- those that give rise to the slowest decaying contributions. Furthermore, the ultimate asymptotic decay is determined by the pole(s) with the smallest imaginary component, referred to as the leading order pole(s). As the model parameters (or statepoint) are varied the identity of the pole with the smallest imaginary component may change, leading to abrupt changes in the type of asymptotic decay. Furthermore, there may also be crossover in oscillatory asymptotic decay with different wavelengths. A number of studies have shown that changes in the asymptotic decay mode may be detected in simulation and experiments and have verified its relevancy in studying the microscopic properties of the fluid~\cite{dijkstra2000ssd,grodon2005hai,baumgartl2005eos,salmon2006dpc,
klapp2008surviving}.

%, as is the case here, the asymptotic, $r>\infty$, of all the total correlation functions, $h_{ij}(r) = g_{ij}(r)-1$, is either exponential (monotonic) or exponentially damped oscillatory. The type of decay, as well as the decay length, and where appropriate, the oscillatory wavelength is the same for all three correlation functions
%with decay length $\alpha_0^{-1}$, or damped oscillatory with decay length $\tilde{\alpha}_0^{-1}$, and oscillatory wavelength $2\pi/\alpha_1$. The type decay is the same for each correlation function, i.e. for $i,j=1,2$~\cite{evans1994adc}. The type of asymptotic decay and the decay length (and oscillatory wavelength) can be determined from the pole structure of, $\hat{h}(q)$, the Fourier transform of $h(r)$ i.e. by determining the values of the complex $q=\alpha_1+\im\alpha_1$ where $\hat{h}(q)$ diverges. The pole with the smallest $\alpha_0$, referred to as the leading order poles, determines the asymptotic decay for all correlation functions. Furthermore, this pole

The {\it raison d'\^etre} of DFT lies in its prowess to investigate inhomogeneous situations in equilibrium. Given an approximation for the density functional, taking the derivative with respect to the density profiles yields a set of Euler-Lagrange equations. By numerically solving these equations one obtains the set of equilibrium density profiles, which minimise the grand potential functional of the system.

In the present paper we consider the NAHS model with $\Delta>0$ and calculate fluid-fluid demixing binodals and spinodals for a range of size ratios, $q$, and non-additivity parameters, $\Delta$. For size ratio $q=0.1$, we compare the predictions for the coexistence curves to results of computer simulations by Dijkstra~\cite{dijkstra1998pbn}. We find that the theory reproduces the location of the binodal reasonably well. By calculating the partial pair direct correlation functions from functional derivatives of the excess free energy functional and inverting the OZ equation in Fourier space, we determine the positions of the poles, as well as the asymptotic, $r\to\infty$, decay of the partial pair correlation functions. We find that besides structural crossover between oscillatory decay with one wavelength to oscillatory decay with a different wavelength (that also occurs in the additive model~\cite{grodon2004dcf}), there is also Fisher-Widom (FW) crossover from exponentially damped oscillatory decay to monotonic exponential decay. We use the functional for investigating the planar fluid-fluid interface between coexisting phases and demonstrate how the different types of asymptotic decay of the bulk correlations in the two coexisting phases determine the asymptotic and intermediate decay of the density profiles on both sides of the fluid-fluid interface, consistent with the general theory of asymptotic decay of correlations~\cite{evans1994adc}. The essential quantity in studying fluid interfaces is the surface tension, $\gamma$, which is the excess free energy per unit area required to maintain the surface. This can be measured experimentally and therefore provides a direct connection between theoretical approaches and real fluids. We determine $\gamma$ quantitatively and show that as the critical point is approached, mean-field scaling is reproduced. 

The paper is structured as follows: In section~\ref{sec:theory}, which can be safely skipped by expert readers, we outline the relevant theory including a description of the excess free energy functional used, the Ornstein-Zernike equation for binary mixtures and the theory of asymptotic decay of correlations. We also describe how to calculate inhomogeneous density profiles. The main results of the work are described in section~\ref{sec:results}, where we present bulk phase diagrams for a range of size ratios and non-additivities. We investigate the pole structure of the structure factors in the complex plane and indicate the regions of the phase diagram with different types of asymptotic decay. We calculate the density profiles for the fluid-fluid interface and the surface tension. In section~\ref{sec:discus} we present a discussion of the results. Finally, in the appendices we present further details on the structure of the weight and kernel functions, and explicitly write down the free energy density contributions.

\section{Theoretical Background}
\label{sec:theory}
\subsection{Density Functional Theory}
We first reintroduce the general density functional theory framework~\cite{evans1992fif,evans1979nlv}. For a classical system composed of two different species of particles, one can construct a grand potential functional $\Omega[\rho_1,\rho_2]$ of the set of one-body density profiles $\rho_i(\rr)$ for $i=1,2$,
\begin{equation}
 \Omega[\rho_1,\rho_2]=F[\rho_1,\rho_2]-\sum_{i=1}^{2}\int\drr\rho_i(\rr)(\mu_i-V\suprm{ext}_i(\rr)),
\label{eq:omega}
\end{equation}
where $F[\rho_1,\rho_2]$ is the intrinsic Helmholtz free energy functional, $\mu_i$ is the chemical potential of species $i$, $V\suprm{ext}_i(\rr)$ is an external potential that acts on particles of species $i$ and $\rr$ is the spatial coordinate. The intrinsic Helmholtz free energy functional may be separated into two contributions: 
\begin{equation}
F[\rho_1,\rho_2]=F_{\mathrm{id}}[\rho_1,\rho_2]+F_{\mathrm{ex}}[\rho_1,\rho_2].
\label{eq:F_sep}
\end{equation}
The first term in~\eref{eq:F_sep} is the Helmholtz free energy of an ideal gas,
\begin{equation}
F_{\mathrm{id}}[\rho_1,\rho_2] = \sum_{i=1}^{2} k_BT \int\drr\rho_i(\rr)(\ln(\Lambda_i^3\rho_i(\rr))-1),
\end{equation}
where $k_BT$ is the thermal energy and $\Lambda_i$ is the thermal de Broglie wavelength of particles of species $i$. The second term in \eref{eq:F_sep}, $F_{\mathrm{ex}}[\rho_1,\rho_2]$, is the excess contribution to the free energy which is due to inter-particle interactions. This part is in general unknown and is specific to the form of the inter-particle interactions. The FMT approximation for the binary NAHS excess free energy functional will be defined below.

% The minimum value of \eref{eq:omega} is equal to the thermodynamic grand potential of the system, i.e.\ $\Omega=\min(\Omega[\rho_1,\rho_2])$, and the set of density profiles which minimize the functional $\Omega[\rho_1,\rho_2]$ are the equilibrium fluid density profiles satisfying the following pair of Euler-Lagrange equations:
% \begin{equation}
% \frac{\delta \Omega[\rho_1,\rho_2]}{\delta\rho_i(\rr)}=0,
% \label{eq:EL_0}
% \end{equation}
% which may be solved to obtain the two equilibrium fluid density profiles for given values of $\mu_1$ and $\mu_2$. 

It can be shown that when $\Omega[\rho_1,\rho_2]$ is minimised w.r.t.~the density distributions, its value is equal to the thermodynamic grand potential of the system, $\bar{\Omega}$, and that the set of density profiles that minimise $\Omega[\rho_1,\rho_2]$ are the set of equilibrium density profiles, $\bar{\rho}_i(\rr)$. One can summarise these two statements as
\begin{equation}
\left.\frac{\delta \Omega[\rho_1,\rho_2]}{\delta\rho_i(\rr)}\right|_
{\bar{\rho}_1,\bar{\rho}_2} = 0,
  \quad \quad \quad  
\Omega[\bar{\rho}_1,\bar{\rho}_2]=\bar{\Omega},
\end{equation}
where the left hand side of the first equation represents the functional derivative of $\Omega[\rho_1,\rho_2]$ with respect to $\rho_i(\rr)$, $i=1,2$, evaluated with the equilibrium density profiles, $\bar{\rho}_1(\rr)$ and $\bar{\rho}_2(\rr)$.

Properties of bulk fluid states can be obtained from evaluating \eref{eq:F_sep} with constant bulk densities, $\rho_i(\rr)=\rho_i^b$, such that the Helmholtz free energy in bulk is
\begin{equation}
\mathcal{F}(\rho_1^b,\rho_2^b) = F[\rho_1^b,\rho_2^b].
\label{eq:bulk_f}
\end{equation}
The chemical potentials and the pressure are given respectively by,
\begin{eqnarray}
\mu_i(\rho^b_1,\rho^b_2)=V^{-1}\frac{\partial \mathcal{F}(\rho_1^b,\rho_2^b)}{\partial \rho^b_i}, \qquad
P(\rho_1,\rho_2)=-\frac{\mathcal{F}(\rho_1^b,\rho_2^b)}{V}+\sum_{i=1}^2\rho^b_i\mu_i,
\label{eq:mu_P}
\end{eqnarray}
where $V$ is the system volume. For a system that exhibits phase separation, coexistence curves (binodals) are obtained by finding pairs of statepoints for which the chemical potentials and the pressure are the same in the two phases, labelled A and B, i.e., by solving simultaneously the three equations;
\begin{eqnarray}
P\suprm{(A)}=P\suprm{(B)} \quad {\rm and} \quad
\mu_i\suprm{(A)}=\mu_i\suprm{(B)},  \, \, i=1,2,
\end{eqnarray}
where $\mu_i\suprm{(A)}$ is the chemical potential of species $i$ in phase A and $P\suprm{(A)}$ is the pressure of phase $\rm{A}$ (and similarly for phase B). The limit of mechanical stability of the system (spinodal) can be obtained from the (numerical) solution of $\det(\partial^2 ( \mathcal{F} / V )/\partial\rho_i \partial\rho_j ) = 0$. At the spinodal the compressibility of the fluid becomes infinite and the correlation length of the fluid, which is the length-scale on which the fluid correlations decay, diverges to infinity. The binodal and the spinodal meet at the critical point, where the correlation length in the coexisting phases becomes infinite and the order parameter that describes the difference between the two phases vanishes, i.e.~the two phases become indistinguishable from each other.

% The {\it total} Helmholtz free energy,dsadsa
% \begin{equation}
% \mathcal{F}[\rho_1,\rho_2]=\mathcal{F}\subrm{id}[\rho_1,\rho_2]+\mathcal{F}\subrm{ex}[\rho_1,\rho_2],
% \label{eq:ftot}
% \end{equation}
% also includes a contribution from the ideal gas,
% \begin{equation}
% \mathcal{F}\subrm{id}[\rho_1,\rho_2] = k_B T  \sum_{i =1,2} \int\dr\rho_i(\rr) [\ln(\rho_i(\rr) \Lambda_i^3 ) - 1]
% \end{equation}
% where $\Lambda_i$ is the (irrelevant) de-Broglie wavelength of species $i$.

\subsection{The Binary Non-Additive Hard Sphere Excess Free Energy Functional}
\label{sec:thefunc}
In FMT the excess Helmholtz free energy functional is constructed from a set of weighted densities, $n_{\nu}^{(i)}(\rr)$, which are formed by convolution of the bare density profiles, $\rho_i(\rr)$, with a set of geometrically inspired weight functions, $w_\nu^{(i)}(\rr)$, appropriate for the model. The index $\nu$ labels the type of weight function. In the binary NAHS functional the weight functions are spherically symmetric so that the weighted densities are given by
\begin{equation}
n_{\nu}^{(i)}(\xx)=\int\dr\rho_i(\rr)w_\nu(|\xx-\rr|,R_i), \quad i=1,2,
\label{eq:wei_dens}
\end{equation}
where $\nu=0, 1, 2, 3$, and $R_i = \sigma_{ii}/2$ is the hard sphere radius of species $i$. The weight functions represent the `fundamental measures' of the model in question, i.e. their volume ($\nu=3$), surface area (2), integral mean curvature (1), and Euler characteristic (0). They have dimension $(\rm{length})^{\nu-3}$, and therefore the weighted densities also have dimension $(\rm{length})^{\nu-3}$. We define components of a free energy density that depend on the weighted densities,
\begin{equation}
\Phi_{\alpha\beta}(\xx,\xx')\equiv\Phi_{\alpha\beta}\left(\{n_{\nu}^{(1)}(\xx)\},\{n_{\tau}^{(2)}(\xx')\}\right),
\label{eq:phi}
\end{equation}
where $\alpha,\beta=0,1,2,3$. These terms have dimension $(\rm{length})^{\alpha+\beta-6}$ and their full form is given below. The excess (over ideal) Helmholtz free energy functional, $F\subrm{ex}[\rho_1,\rho_2]$, is given by a double integral over space and double sum over the geometric indices,
\begin{equation}
\frac{F\subrm{ex}[\rho_1,\rho_2]}{k_BT}=\sum_{\alpha,\beta=0}^3\int\int\dx\dx'  \Phi_{\alpha\beta}(\xx,\xx')K_{\alpha\beta}(|\xx-\xx'|),
\label{eq:Fex}
\end{equation}
where the convolution kernels, $K_{\alpha\beta}(r)$, control the range of non-additivity between unlike components. 
The kernel functions are isotropic and are similar to the weight functions, although they depend on a new length scale, 
\begin{equation}
R_{12}=\Delta(R_{1}+R_{2})=\sigma_{12}-\tfrac{1}{2}(\sigma_{11}+\sigma_{22}),
\label{eq:R12}
\end{equation}
which is the difference between the cross-species diameter and the mean particle diameter. (Note that in general $R_{12}\neq\sigma_{12}/2$). The kernel functions $K_{\alpha\beta}$ have dimension $(\rm{length})^{-\alpha-\beta}$, therefore the products $\Phi_{\alpha\beta}K_{\alpha\beta}$ have the correct dimension $(\rm{length})^{-6}$, as required by~\eref{eq:Fex}.

We use the (fully scalar) Kierlik-Rosinberg form for $w_\nu(r, R )$. Hence the four weight functions used in Eq.~\eref{eq:wei_dens} are defined as
\begin{eqnarray}
w_3(r,R) &=\sgn(R)\Theta( R-r ), \nonumber \\
w_2(r,R) &=\delta( R-r ), \label{eq:wei_fncs} \\
w_1(r,R) &=\tfrac{\sgn(R)}{8\pi}\delta' ( R-r ), \nonumber \\
w_0(r,R) & = -\tfrac{1}{8\pi}\delta''( R-r ) + \tfrac{1}{2\pi r}\delta' ( R-r ), \nonumber
\end{eqnarray}
where $r=|\rr|$, $R = R_1,R_2$, $R_{12}$, $\sgn(\cdot)$ is the sign function, $\Theta(\cdot)$ is the Heaviside step function, $\delta(\cdot)$ is the Dirac distribution, and the prime denotes the derivative w.r.t.~the argument.  Although the hard sphere radii $R_i$ are strictly greater than zero, the factor $\sgn(R)$ is included in order that the weight functions may be reused below for the kernel functions, where $R_{12}$ may be negative (if $\Delta<0$).

The set of convolution kernels are symmetric w.r.t. exchange of indices, $K_{\alpha\beta}=K_{\beta\alpha}$, so that there are only ten independent weight functions. Four of these are given via~\eref{eq:wei_fncs} with $R = R_{12}$;
\begin{eqnarray}
\eqalign
K\!_{00}(r) &= w_3(r,R_{12}), \qquad K_{01}(r)&=w_2(r,R_{12}), \label{eq:wei_fncs2}\\
K_{02}(r) &= w_1(r,R_{12}), \qquad K_{03}(r)&=w_0(r,R_{12}). \nonumber%
\end{eqnarray}
The set of further weight functions, suppressing for notational convenience the arguments of $w_\nu^{(\dag)}(r,R_{12})$, are: 
\begin{eqnarray}
K_{11}(r) = w^\dag_1 \;\;=&\sgn(R)\delta'( R-r ),& \nonumber\\
K_{12}(r)=w_0^\dag \;\;= &\tfrac{1}{8\pi}\delta''( R-r ),&\nonumber\\ 
K_{22}(r)= w_{-1}^\dag\;\;= &\tfrac{1}{64\pi^2}\delta^{(3)} ( R-r ), & \nonumber\\
K_{13}(r)=w_{-1}=&\sgn(R)\left[\tfrac{1}{2\pi r}\delta''( R-r )-\tfrac{1}{8\pi}\delta ^{(3)} ( R-r )\right],& \\
K_{23}(r)= w_{-2}   = &\tfrac{1}{16\pi^2r}\delta^{(3)} ( R-r )-\tfrac{1}{64\pi^2}\delta^{(4)} ( R-r ),& \nonumber \\ 
K_{33}(r) =  w_{-3} =&\tfrac{\sgn(R)}{8\pi^2}\left[−\tfrac{1}{r}\delta^{(4)} ( R-r ) + \tfrac{1}{8}\delta^{(5)} ( R-r )\right], & \nonumber
\label{eq:k_terms}
\end{eqnarray}
where $R=R_{12}$, and the derivatives of the Dirac delta function are defined by $\delta^{(\gamma)}(x ) =\dd^\gamma \delta(x )/\dd x^{\gamma}$ for $\gamma = 3, 4, 5$. The Fourier space expressions of all weight functions are given explicitly in~\ref{sec:weifncs_fs}.

The terms $\Phi_{\alpha\beta}$ are built from a sum of derivatives of the zero-dimensional excess free energy, $\phi \subrm{0d}(\eta) = (1-\eta)\ln(1-\eta) + \eta$, where $\eta$ is a dummy argument (which can be viewed as the average occupation number of a zero dimensional cavity~\cite{rosenfeld1997fundamental}), and $\gamma$ labels the derivative: $\phi\subrm{0d}^{(\gamma)}(\eta)\equiv \dd^\gamma \phi\subrm{0d}(\eta)/\dd \eta^\gamma$. The derivatives of $\phi \subrm{0d}(\eta)$ are multiplied by products of weighted densities to ensure the correct dimensionality of the free energy density. We introduce ansatz functions $A_{\alpha\gamma}^{(i)}$ that possess the dimension of $(\rm{length})^{\alpha-3}$ and the order $\gamma$ in density (i.e. they contain $\gamma$ factors $n_{\nu}^{(i)}$). These are combined as
\begin{equation}
\Phi_{\alpha\beta}=\sum_{\gamma=0}^6\sum_{\gamma'=0}^3A_{\alpha\gamma'}^{(1)}A_{\beta(\gamma-\gamma')}^{(2)}\phi\subrm{0d}^{(\gamma)}\left(n_3^{(1)}+n_3^{(2)}\right).
\label{eq:phi2}
\end{equation}
Expressions for the non-vanishing terms of the ansatz functions are,
\begin{equation}
\label{eq:prefac}
\begin{array}{llll}
A_{01}^{(i)}=n_0^{(i)},\qquad& A_{02}^{(i)} = n_1^{(i)}n_2^{(i)}, \qquad&A_{03}^{(i)}= \tfrac{1}{2\pi}(n_2^{(i)})^3, & \\
A_{11}^{(i)}=   n_1^{ (i )},& A_{12}^{(i )}=\tfrac{1}{8\pi}(n_{2}^{(i )} )^2 , &	 
A_{21}^{(i )}=     n_2^{(i )}, &A_{30}^{(i )}=1. 
\end{array}
\end{equation}
% \begin{eqnarray}
%  A_{01}^{(i)}&=&n_0^{(i)},& A_{02}^{(i)}& = n_1^{(i)}n_2^{(i)}, \\
%   A_{03}^{(i)}&=& (n_2^{(i)})^3 /(2\pi),& A_{11}^{(i)}&=    n_1^{ (i )}        \\
%  A_{12}^{(i )}&=(n_{2}^{(i )} )^2 /(8\pi),	&          A_{21}^{(i )}&=     n_2^{(i )} , \\
%  A_{30}^{(i )}&=1
% \end{eqnarray}
The specific form of~\eref{eq:prefac} ensures both that the terms in the sum in~\eref{eq:Fex} possess the correct dimension of $(\rm{length})^{-6}$ and that the prefactor of $\phi\subrm{0d}$ in~\eref{eq:phi2} is of the total order $\gamma$ in densities.

This completes the prescription for the excess Helmholtz free energy functional. Evaluating the sums in Eq.~\eref{eq:phi2} explicitly results in a total of 49 terms, which can be grouped either by the 16 kernel functions, or alternatively by the 10 unique weight functions. In \ref{sec:phi_terms_apx} we transcribe some of the terms; all further terms can be obtained by symmetry.

\subsection{Fluid Structure and Asymptotic Decay of Correlations}
\label{sec:asymp_decay}
In order to study the pair structure of the bulk fluid, rely on the OZ equation, which separates the partial pair distribution functions, $g_{ij}(r)$, into a `direct' part between pairs of particles, and an `indirect' part that comes from the interaction between all the other particles in the system:
\begin{equation}
h_{ij}(r)=c_{ij}(r)+\sum_{l=1}^2\rho_l^b\int\dd \rr' h_{il}(r') c_{lj}(|\rr-\rr'|),
\label{eq:OZ}
\end{equation}
where $h_{ij}(r)=g_{ij}(r)-1$ is the total correlation function and $c_{ij}(r)$ is the two-body direct correlation function between species $i$ and $j$. The latter can be obtained from the excess free energy functional via functional differentiation,
\begin{equation}
c_{ij}(|\rr-\rr'|)=-(k_BT)^{-1}\left.\frac{\delta^2 F\subrm{ex}}{\delta\rho_i(\rr)\delta\rho_j(\rr')}\right|_{\rho_1,\rho_2=\rm{const}}.
\label{eq:c2}
\end{equation}
By Fourier transforming one can re-write~\eref{eq:OZ} as
\begin{equation}
\hat{h}_{ij}(k)=\hat{c}_{ij}(k)+\sum_{l=1}^2\rho_l^b \hat{h}_{il}(k)\hat{c}_{lj}(k),
\label{eq:OZ2}
\end{equation}
where $\hat{h}_{ij}(k)$ is the (three-dimensional) Fourier transform of $h_{ij}(r)$,
\begin{equation}
\hat{h}_{ij}(k)=\frac{4\pi}{k}\int_0^\infty\dd r \, r\sin(kr)h_{ij}(r),
\label{eq:ft}
\end{equation}
and similarly for $\hat{c}_{ij}(k)$. It can be shown by rearranging the OZ equations that
\begin{eqnarray}
\hat{h}_{ij}(k)=\frac{\hat{N}_{ij}(k)}{\hat{D}(k)},
\label{eq:heqnod}
\end{eqnarray}
where the common denominator is
\begin{equation}
\hat{D}(k)=[1-\rho_1\hat{c}_{11}(k)][1-\rho_2\hat{c}_{22}(k)]-\rho_1\rho_2\hat{c}_{12}(k),
\label{eq:denom}
\end{equation}  
and the numerators in~\eref{eq:heqnod} depend on the species indices:          
 \begin{eqnarray}
\hat{N}_{11}(k)&= \hat{c}_{11}(k) + \rho_2[\hat{c}_{12}^2(k) - \hat{c}_{11}( k) \hat{c}_{22} (k)], \nonumber \\
\hat{N}_{22}(k)&= \hat{c}_{22}(k) + \rho_1[\hat{c}_{12}^2(k) - \hat{c}_{11}( k) \hat{c}_{22} (k)] , \label{eq:numers} \\
\hat{N}_{12}(k)&=\hat{c}_{12}(k).  \nonumber 
 \end{eqnarray}
Using the definition of the inverse Fourier Transform, we obtain
\begin{eqnarray}
h_{ij}(r)&=&\frac{1}{2\pi^2r}\int_0^\infty\dd k\,k\sin(kr)\hat{h}_{ij}(k), \nonumber \\
&=&\frac{1}{2\pi^2r}\int_0^\infty\dd k \, k\sin(kr)\frac{\hat{N}_{ij}(k)}{\hat{D}(k)}.
\label{eq:h_kint}
\end{eqnarray}
For the present functional, expressions for $\hat{c}_{ij}(k)$ can be obtained analytically via~\eref{eq:c2} and hence can be substituted into~\eref{eq:denom} and \eref{eq:numers}, before numerically Fourier transforming to obtain $g_{ij}(r)=h_{ij}(r)+1$ from \eref{eq:h_kint}. No numerical scheme for solving~\eref{eq:OZ} is required. Indeed this method has already been successfully used in~\cite{schmidt2004rfn} and \cite{schmidt2007fmd} to calculate the distribution functions and partial structure factors,
\begin{equation}
S_{ij}(k)=\delta_{ij}+\sqrt{\rho_1^b\rho_2^b}\hat{h}_{ij}(k),
\label{eq:sk}
\end{equation}
where $\delta_{ij}$ is the Kronecker delta.
%Inverting the OZ relations permits us to calculate partial structure factors, $S_{ij}(k )$, and partial pair correlation functions, $g_{ij}(r)$. Furthermore, b

Another method that we will make extensive use of in the following is to investigate the singularities of $\hat{h}_{ij}(k)$ in the complex $k$-plane~\cite{evans1994adc}. Using~\eref{eq:h_kint} and assuming that the singularities of $\hat{h}_{ij}(k)$ for the present systems are simple poles, we can proceed via Cauchy's residue theorem. Performing contour integration around a semicircle in the upper half of the complex $k$-plane, the total correlation functions can be written as a sum of contributions from the poles enclosed,
\begin{equation}
rh_{ij}(r) = \sum_n A_n^{(ij)} \exp(\im k_nr),
\end{equation}
where $n$ labels the poles, $k_n$ satisfies $\hat{D}(k_n) = 0$, $A_n^{(ij)}$ is the amplitude associated with the pole at $k_n$ and $\im$ is the imaginary unit. The amplitude is related to the residue $R_n^{(ij)}$ by $A_n^{(ij)} = R_n^{(ij)}/2$.

The poles are either purely imaginary, $k_n = \im\alpha_0$, or occur as a complex pair, $k_n= \pm \alpha_1 + \im\alpha_0$, where both $\alpha_0$ and $\alpha_1$ are real. In general there will be an infinite number of poles and contributions from many of those are required to account for the behaviour of $h_{ij}(r)$ at small distances $r$. However, the ultimate, $r\to\infty$, decay of all $h_{ij}(r)$ is determined by the pole(s) that gives the slowest exponential decay, i.e., the pole(s) with the smallest imaginary part $\alpha_0$. These are referred to as the leading order pole (or poles in the case of a conjugate complex pair).

If the leading order pole is purely imaginary, then all $rh_{ij}(r)$ ultimately decay exponentially, $rh_{ij}\sim A_{ij} \exp(-\alpha_0 r)$, as $r\to\infty$, where $A_{ij}$ is an amplitude specific to each correlation function. On the other hand, if the leading order poles are a conjugate pair, then the sum of contributions from this pair of complex poles gives damped oscillatory ultimate decay, $rh_{ij}(r)\sim 2A_{ij} \exp( -\alpha_0r) \cos(\alpha_1 r - \theta_{ij})$, with a common characteristic decay length $\alpha_0^{-1}$ and wavelength of oscillations $2\pi/\alpha_1$. $A_{ij}$ and $\theta_{ij}$ denote the amplitude and the phase, respectively.

As the model parameters and statepoint change, the positions of the poles in the complex plane vary. The pole(s) which have the smallest imaginary part, referred to as the leading order pole(s), can therefore be replaced by a different set of poles. This can lead to abrupt changes in the type of decay, either between damped oscillatory decay and monotonic decay, or between damped oscillatory decay with one wavelength to damped oscillatory with a different wavelength.

For the one-component hard sphere fluid the decay is always damped oscillatory with a wavelength similar to the hard sphere diameter~\cite{evans1994adc}. For the binary additive hard sphere mixture there is an abrupt crossover in the phase diagram from one wavelength similar to the diameter of species 1 to a different wavelength similar to the diameter of species 2~\cite{grodon2004dcf}. The two different oscillatory wavelengths are each described by a complex conjugate pair of poles with real components which determine the oscillatory wavelength. This abrupt change occurs when these two pairs of poles have the same imaginary component. This marks a {\it structural crossover line} in the phase diagram.

In general, there may also be a crossover between damped oscillatory and monotonic decay, particularly in systems which exhibit phase separation and where correlation functions obey Ornstein-Zernike (asymptotic exponential decay) behaviour close to the critical point. This crossover can occur via two mechanisms: In {\it Fisher-Widom crossover} a pair of leading order complex poles and a single imaginary pole change their positions in the complex plane as the statepoint is varied. As the critical point is approached the leading order pole(s) change from the complex pair to the purely imaginary pole. The statepoints where this crossover occurs traces the FW line in the phase diagram.

It has been shown, both via simulation and theory, that additive mixtures of hard spheres with small size-ratio, $q\lesssim0.2$, can exhibit (meta-stable) fluid-fluid phase separation~\cite{PhysRevE.59.5744}. However, the PY approximation is unable to account for this phenomenon. Correspondingly, the asymptotic decay of the distribution functions is always oscillatory, for all size ratios, within PY theory~\cite{grodon2004dcf}. In the same study, the authors also consider an effective one-component depletion potential, from which they are able to obtain the phase transition and to also find Fisher-Widom crossover form oscillatory to monotonic decay. Since the NAHS functional, taken in the additive limit $\Delta = 0$, recovers the PY approximation for the bulk correlation functions, we do not see Fisher-Widom crossover in the additive model.

In {\it Kirkwood crossover} two purely imaginary poles come together, coalesce and become a pair of complex poles. This mechanism often occurs in fluids that interact via soft, steeply-repulsive, pair potentials~\cite{decarvalho1999dcf,hopkins2005adp}. Indeed we do not find it in the present system.

% As the statepoint is varied two purely imaginary poles move together, coalesce and then become a pair of complex poles. This crossover mechanism often occurs in fluids where the particles interact via soft steeply-repulsive pair potentials, e.g.~\cite{hopkins2005adp,decarvalho1999dcf}.

%This region of exponential decay bounding the spinodal is consistent with the results from other phase separating models, e.g.~\cite{archer2001bgc,hopkins2006pcf}.
%Note that whereas the wavelength $2\pi/\alpha_1$ and the decay lengths $\alpha_0^{-1}$ or $\tilde{\alpha}_0^{-1}$ are the same for all $h_{ij}(r)$, the amplitudes and phases do depend on the indices $ij$. However, general considerations demand $A_{12} = A_{11}A_{22}$ or $\tilde{A}_{12} = \tilde{A}_{11}\tilde{A}_{22}$.

\subsection{Inhomogeneous Systems}
\label{sec:inhomo_systems}
In order to calculate inhomogeneous density profiles in the grand-canonical ensemble we consider the thermodynamic grand potential functional~\eref{eq:omega} and minimise $\Omega[\rho_1,\rho_2]$ with respect to variations in the density profiles. This is equivalent to solving a pair of Euler-Lagrange equations,
\begin{equation}
\mu_i= k_BT\log(\Lambda^3\rho_i)-k_BT c^{(1)}_i(\rr) + V_i^{\mathrm{ext}}(\rr), \qquad i=1,2, %\frac{\delta F}{\delta\rho_i(\rr)}+V_i(\rr),
\label{eq:EL_0}
\end{equation}
where $c^{(1)}_i(\rr)=-(k_BT)^{-1} \delta F_{\rm ex}/\delta\rho_i(\rr)$ is the one-body direct correlation functional for species $i=1,2$. In practice, the pair of equations~\eref{eq:EL_0} must be solved simultaneously via an iterative numerical scheme. Explicit functional derivation of the excess part of the present functional, \eref{eq:Fex}, yields
\begin{eqnarray}
\frac{\delta \mathcal{F}\subrm{ex}}{\delta \rho_i(\rr)}&=\sum_{\gamma=0}^3\int\dd \xx w_\gamma^{(i)}(|\xx-\rr|) \nonumber \\
& \quad \quad \left[ \sum_{\alpha,\beta=0}^3	\int \dd \xx' \frac{\partial \Phi_{\alpha\beta}}{\partial n_\gamma^{(i)}}(\xx,\xx')K_{\alpha\beta}(|\xx-\xx'|)\right],
\label{eq:c1_term}
%\sum_{\alpha,\beta,\gamma=0}^3
\end{eqnarray}
%where $\alpha,\beta$ and $\gamma$ run from 0 to 3.
which has the structure of two nested convolutions. For each value of $\gamma$, the partial derivatives of the excess free energy terms, $\phi_{\alpha\beta\gamma}=\partial \Phi_{\alpha\beta} / \partial n_\gamma^{(i)}$, are first convolved with the kernel functions, $K_{\alpha\beta}(r)$, and then the sum of these is convolved with the single-particle weight function, $w_\gamma(r)$. There are a total of 60 terms of the form of \eref{eq:c1_term} to be evaluated.

In the present study we consider the planar fluid-fluid interface between co-existing fluid phases, where $V\suprm{ext}_i(\rr)=0$. The boundary conditions are chosen so that the density profiles decay to the coexisting bulk values far away from the interface.% We use the coexisting statepoints calculated in the previous section to study the one-body density profiles, $\rho_i(z)$, for the planar free interface between the two coexisting phases. By varying the statepoint along the binodal so that we cross the Fisher-Widom line we are also able to demonstrate the effect that the asymptotic decay mode has on the macroscopic density profiles.

\section{Results}
\label{sec:results}

\subsection{Fluid Demixing Phase Diagram}
\label{sec:res_pd}

\begin{figure}[t]
\centering
\includegraphics[width=8.5cm]{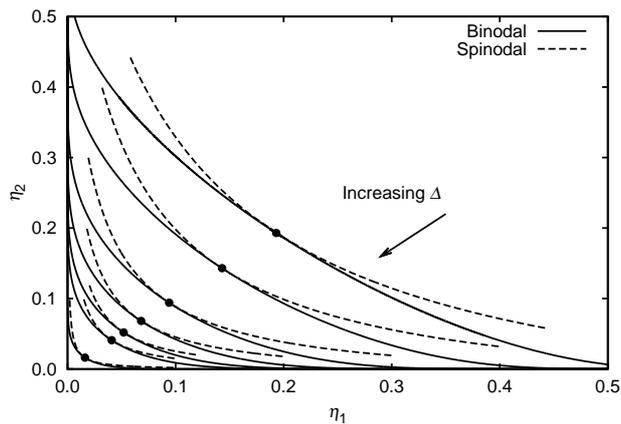}
\caption[Phase Diagram, q=1.0, $\delta=0.1$.]{\label{fig:pd_q1.0_d-.-}
Bulk fluid-fluid binodals (solid lines) and spinodals (dashed lines) for the binary non-additive hard sphere fluid with fixed size ratio $q=\sigma_{11}/\sigma_{22}=1$ and varying non-additivity parameter $\Delta=0.05$, 0.1, 0.2, 0.3, 0.4, 0.5 and 1 (from top to bottom), plotted as a function of the partial packing fractions, $\eta_1$ and $\eta_2$. For each system the binodal meets the spinodal at the bulk critical point ($\bullet$).} 
\end{figure}

\begin{figure}[t]
\centering
\includegraphics[width=8.5cm]{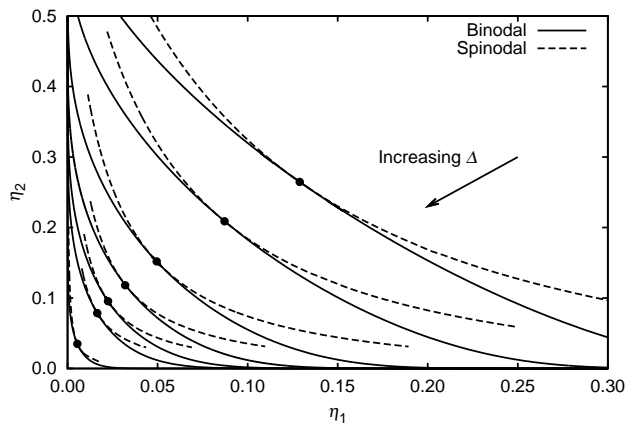}
\caption[Phase Diagram, q=2.0, $\delta=0.1$.]{\label{fig:pd_q2.0_d-.-}
Same as figure~\ref{fig:pd_q1.0_d-.-}, but for fixed size ratio $q=\sigma_{11}/\sigma_{22}=0.5$ and varying $\Delta=$ 0.05, 0.1, 0.2, 0.3, 0.4 and 1 (from top to bottom).} 
\end{figure}

\begin{figure}[t]
\centering
\includegraphics[width=8.5cm]{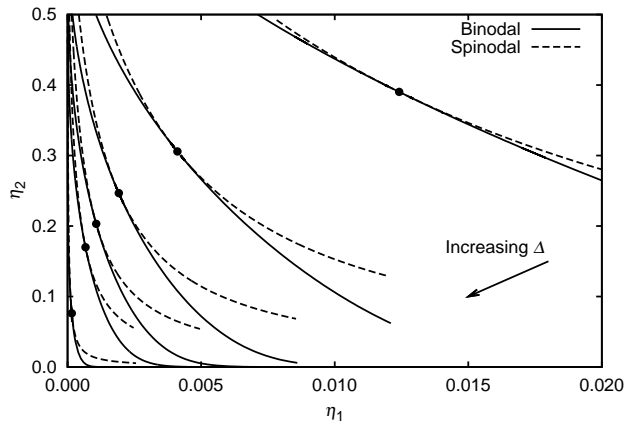}
\caption[Phase Diagram, q=2.0, $\delta=0.1$.]{\label{fig:pd_q10.0_d-.-}
Same as figure~\ref{fig:pd_q1.0_d-.-}, but for fixed size ratio $q=0.1$ and varying $\Delta=$ 0.1, 0.2, 0.3, 0.4, 0.5 and 1 (from top to bottom). Note the scale on the (horizontal) $\eta_1$-axis.} 
\end{figure}

It has previously been shown~\cite{schmidt2004rfn} that for suitable parameters the present theory reproduces the phenomenon that the mixture separates into two different fluid phases~\cite{frenkel1994sep}. Using~\eref{eq:mu_P} and solving for coexisting states we have calculated the coexistence curves (binodals), as well as the spinodals, for a range of model parameters. The statepoint is specified by the partial packing fractions, $\eta_i=\pi\rho_i^b\sigma_{ii}^3/6$; recall that $\rho_i^b$ is the bulk number density for species $i$.

Figure~\ref{fig:pd_q1.0_d-.-} shows the binodals and spinodals for the symmetric mixture, $q=\sigma_{11}/\sigma_{22}=1$, with $\Delta$ varying between 0.05 and 1, in the ($\eta_1$,$\eta_2$) plane. Increasing $\Delta$ causes phase separation at increasingly lower densities, and therefore the partial packing fractions at the critical point, $\eta\suprm{crit}_i$, both decrease monotonically as $\Delta$ increases. Figure~\ref{fig:pd_q2.0_d-.-} displays the binodals and spinodals for asymmetric mixtures with fixed $q=0.5$ and $\Delta$ again varying between 0.05 and 1. On increasing $\Delta$, again both $\eta\suprm{crit}_1$ and $\eta\suprm{crit}_2$ decrease monotonically.

We also consider a mixture with large size asymmetry, $q=0.1$, where we can compare to Gibbs ensemble Monte Carlo simulation results of Dijkstra~\cite{dijkstra1998pbn}. This large asymmetry is a significant test for the functional, as previous studies have showed that FMT struggles with large size asymmetry already in the additive case~\cite{herring2006simulation,herring2007hard}. Figure~\ref{fig:pd_q10.0_d-.-} shows the binodals and spinodals for $q=0.1$ and $\Delta$ varying between 0.1 and 1. Although we show the binodal for $\Delta=0.1$, in simulations it was found that fluid-fluid phase separation for this value of $\Delta$ is metastable with respect to crystallisation \cite{dijkstra1998pbn}. Note that the relevant range of values of $\eta_1$ is much smaller than in figures~\ref{fig:pd_q1.0_d-.-} and~\ref{fig:pd_q2.0_d-.-}. 

\begin{figure}[t]
\centering
\includegraphics[width=8.5cm]{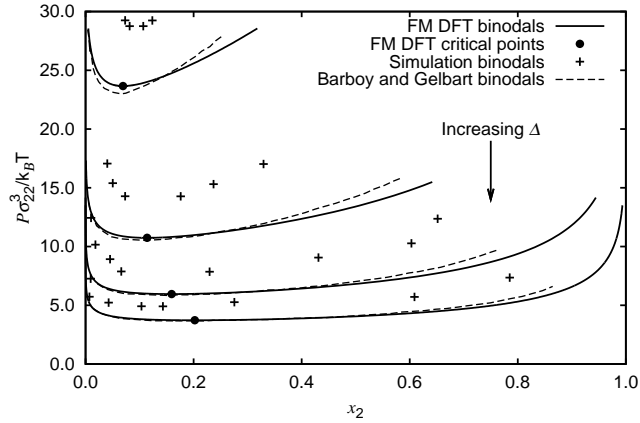}
\caption[Phase Diagram, q=2.0, $\delta=0.1$.]{\label{fig:pd_q10.0_d0.1}
Fluid-fluid coexistence curves for fixed size ratio $q=0.1$ and varying $\Delta=$ 0.2, 0.3, 0.4 and 0.5, plotted in the plane of pressure, $P$, and relative concentration of the large particles $x_2=\rho_2/\rho$. Results from present density functional (solid lines) are compared to the results from Gibbs ensemble simulations (crosses), as reported in~\cite{dijkstra1998pbn}, and a mean-field theory due to Barboy and Gelbart~\cite{barboy1979sre} (dashed lines). The pressure is scaled by the size of the larger species so that the figure is consistent with figure~3 of~\cite{dijkstra1998pbn}. Both theories predict phase separation at lower pressures than the simulation results.}
\end{figure}

To compare our results to those of~\cite{dijkstra1998pbn}, in figure~\ref{fig:pd_q10.0_d0.1} we plot the binodals in the plane spanned by pressure, $P$, and  relative concentration of the larger particles, $x_2=\rho_2/\rho$, alongside the simulation results of Dijkstra and the results from Barboy and Gelbart's mean-field theory~\cite{barboy1979sre} (data taken from~\cite{dijkstra1998pbn}). We find that the coexistence curves from the three approaches have similar shapes and positions, for all values of $\Delta$ shown. However, both theories predict demixing pressures that are lower than the simulation results. Such a systematic error is often a feature of mean-field theories, which underestimate the strength of density fluctuations close to the critical point.

\subsection{Asymptotic Decay of Correlations}

\begin{figure}[t]
\centering
\includegraphics[width=8.5cm]{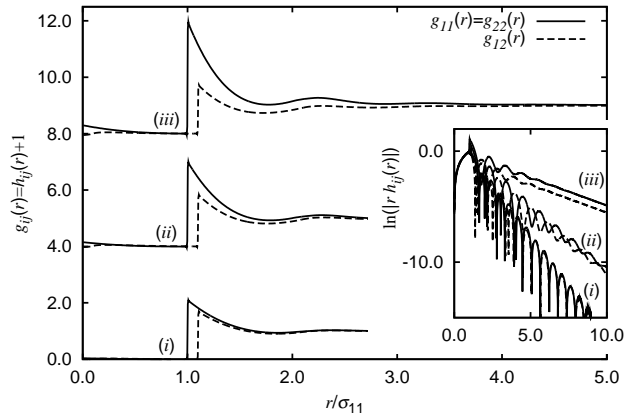}
\caption[Rdfs, q=1.]{\label{fig:g_r_q1}
The partial radial distribution functions, $g_{ij}(r)$, for the system with $q=1$ and $\Delta=0.1$. Note that $g_{11}(r)=g_{22}(r)$ and that the pairs of curves are offset upwards by 4 units for clarity. The pairs of profiles are at the following statepoints: ($i$) $\eta_1=\eta_2=0.1$, ($ii$) $\eta_1=\eta_2=0.13$, and ($iii$) $\eta_1=\eta_2=0.14$. As the density increases and the statepoint approaches the binodal, the oscillations in all $g_{ij}(r)$ become more pronounced. The inset shows $\ln|r\,h_{ij}(r)|$ where $h_{ij}(r)=g_{ij}(r)-1$ is the total correlation function. For ($i$) the intermediate and asymptotic decay is oscillatory with a wavelength $\sim\sigma_{11}$. For ($ii$) the decay is oscillatory at small $r$, but as distance increases, the relative amplitude of the oscillations quickly decreases, and the profiles start to decay monotonically. For ($iii$) there is some short-range oscillations which die out by $r\simeq7\sigma_{11}$ so that the intermediate (and asymptotic) decay is monotonic.}
 \end{figure}

\begin{figure}[t]
\centering
\includegraphics[width=8.5cm]{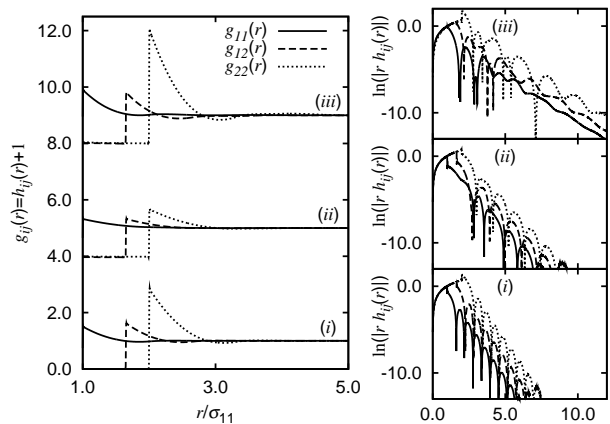}
\caption[Rdfs, q=2.]{\label{fig:g_r_q2}
Same as figure~\ref{fig:g_r_q1}, but for $q=0.5$ and $\Delta=0.1$. The sets of profiles correspond to the following statepoints ($i$) $\eta_1=0.151$, $\eta_2=0.001$, ($ii$) $\eta_1=0.051$, $\eta_2=0.051$, and ($iii$) $\eta_1=0.151$, $\eta_2=0.07$. In the three subplots we display $\ln|r\,h_{ij}(r)|$ for the three different statepoints. In ($i$) the intermediate and asymptotic decay is oscillatory with a wavelength $\sim\sigma_{11}$. In ($ii$) the intermediate and asymptotic decay is oscillatory is with a wavelength $\sim\sigma_{22}=2\sigma_{11}$. In ($iii$) the intermediate decay is oscillatory, but the asymptotic decay is monotonic.}
 \end{figure}

\begin{figure}[t]
\centering
\includegraphics[width=8.5cm,height=5.5cm]{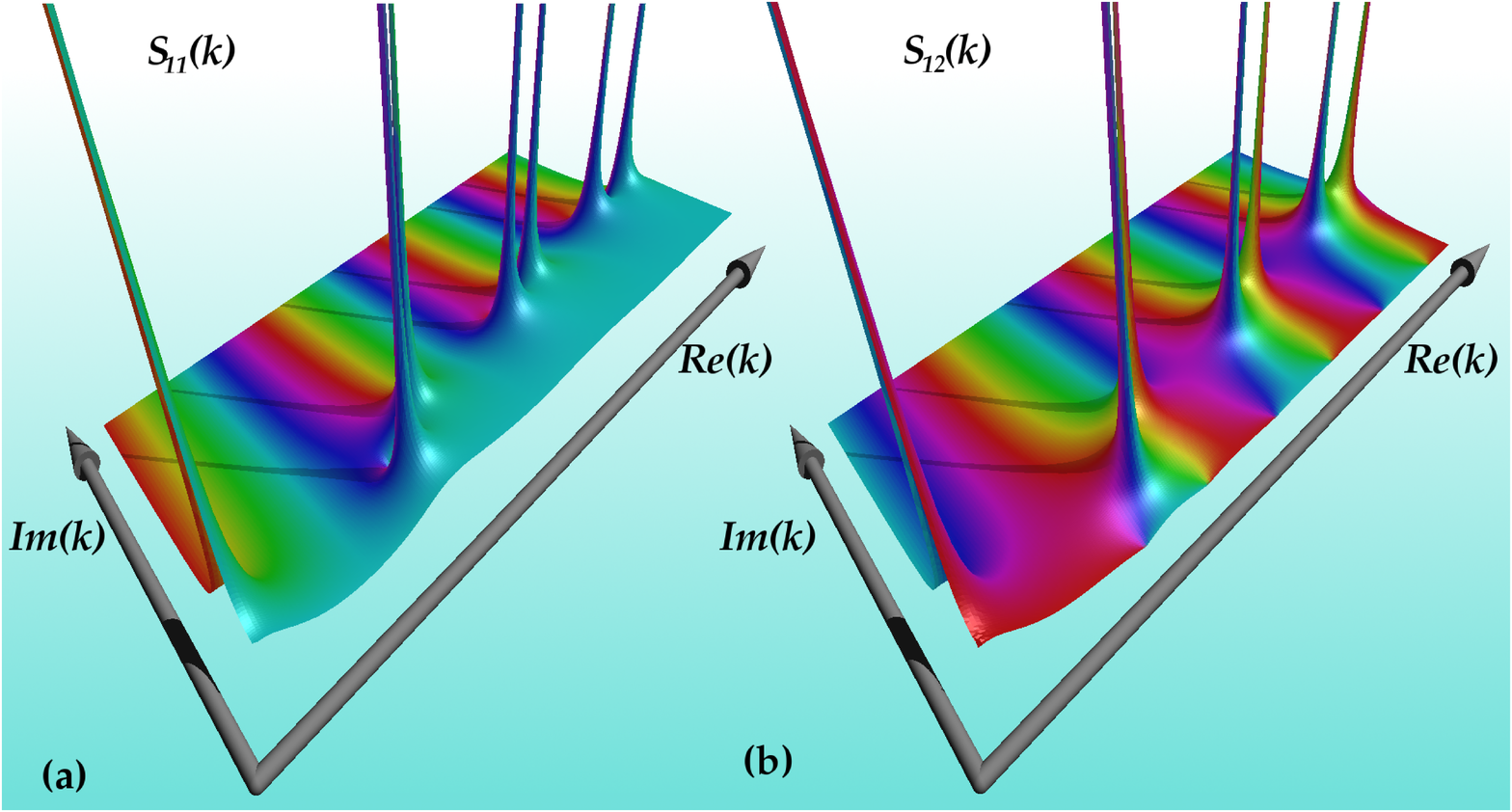}\\
\includegraphics[width=4.25cm]{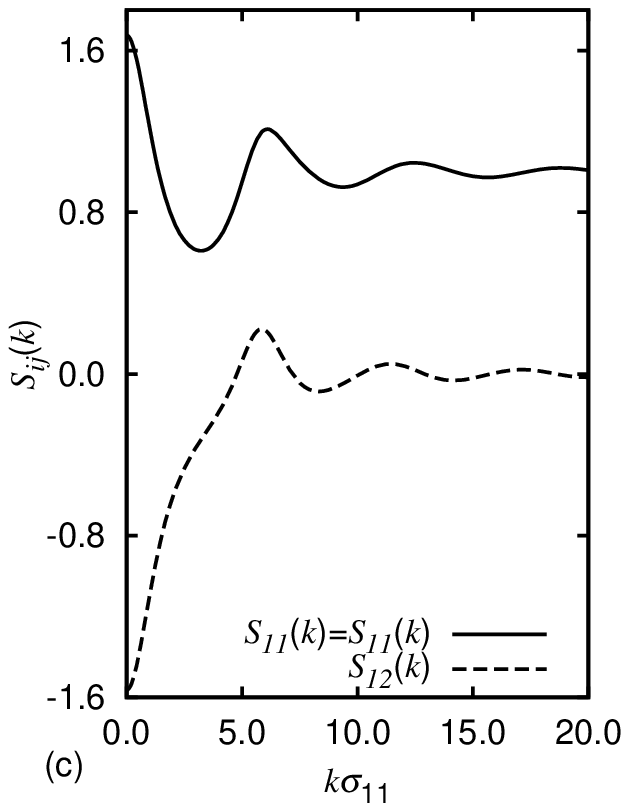}
\caption[3D Structure Factors.]{\label{fig:cmplx_strc_fctr}
Parts (a) and (b): Complex partial structure factors, $S_{ij}(k)$, where $k$ is the complex wave-number, for the symmetric mixture, $q=1$, with $\Delta=0.1$, at statepoint $\eta_1=\eta_2=0.12$. The height and colour of the surfaces represents the absolute value and the complex argument, respectively, of $S_{ij}(k)$. The surfaces are plotted over the range $k=0$ to $k=20+10\im$. Since the mixture is symmetric the intra-species structure factors are identical, i.e $S_{11}(k)=S_{22}(k)$. Note that the two structure factors both have strong divergences (poles) at identical positions. Part (c): The real partial structure factors $S_{ij}(k)$ plotted along the real axis, at the same statepoint as (a) and (b).}
 \end{figure}

Despite the presence of subtle artifacts~\cite{ayadim2010generalization} in the results for the radial distribution functions, $g_{ij}(r)$, calculated via the test particle route from the present functional, it has previously been shown that there is good agreement between the $g_{ij}(r)$, calculated via the Ornstein-Zernike route and Monte Carlo simulation data, both in 3D~\cite{schmidt2004rfn}, and in 1D~\cite{schmidt2007fmd}. Here we explore in detail the asymptotic, $r\to\infty$, decay of correlations which is determined by the poles of the partial structure factors, $S_{ij}(k)$, in the complex plane. We expect the asymptotic decay to be robust and not be affected by the test particle artifacts.

We begin by showing representative examples of the radial distribution functions obtained from numerically Fourier transforming the analytical expressions for $h_{ij}(k)$~\eref{eq:h_kint}. For the symmetric mixture, $q=1$ and $\Delta>0$, by varying the statepoint and calculating $g_{ij}(r)$, we find that there are two types of intermediate and asymptotic, $r\to\infty$, decay. In figure~\ref{fig:g_r_q1} we plot $g_{ij}(r)$ for the symmetric mixture, $q=1$, with $\Delta=0.1$, at three statepoints ($i$) $\eta_1=\eta_2=0.1$, ($ii$) $\eta_1=\eta_2=0.13$, and ($iii$) $\eta_1=\eta_2=0.14$. We find that as the statepoint approaches the binodal, the oscillations in $g_{ij}(r)$ become more pronounced, when viewed on a linear scale.

 \begin{figure}[t]
\centering
\includegraphics[width=8.5cm]{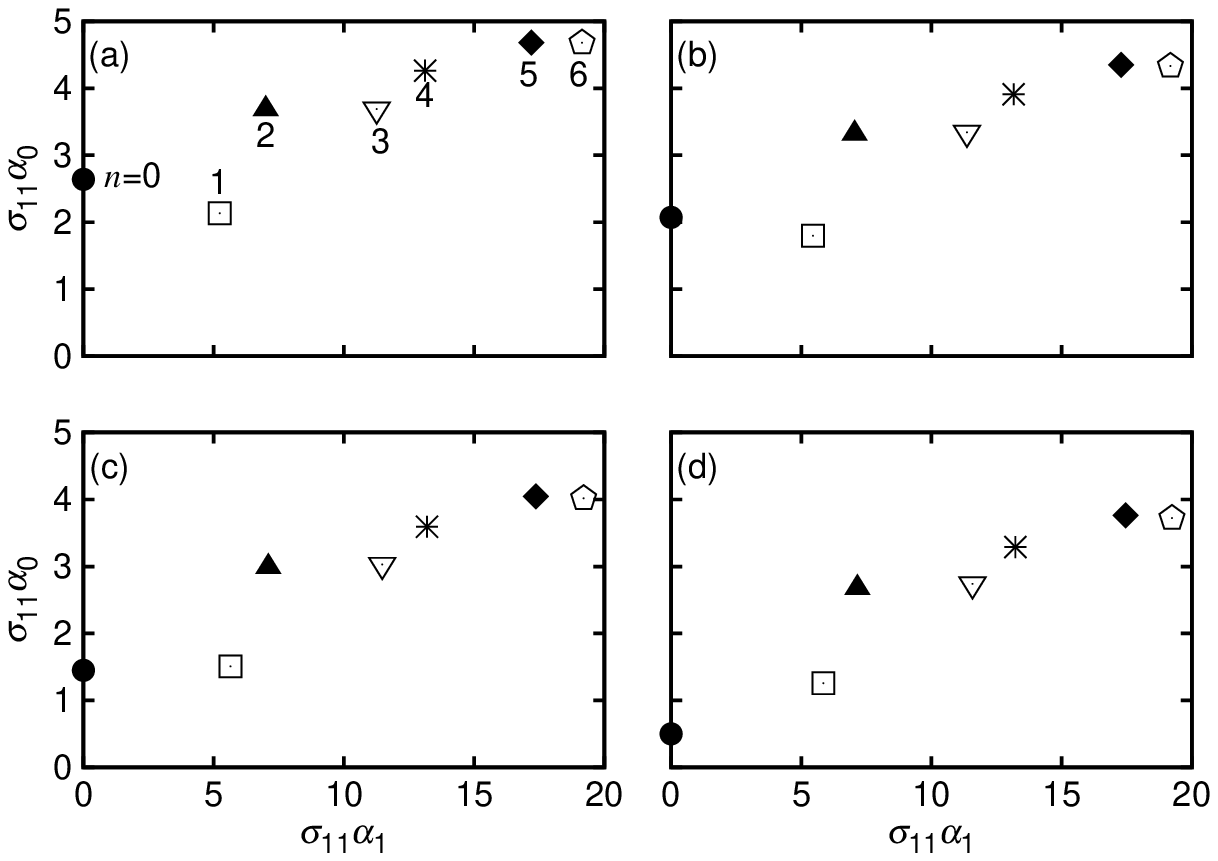}
\caption[Phase Diagram, $q=1.0$, $\Delta=0.1$.]{\label{fig:poles_R1.0}
Positions of the complex poles, $k_n=\alpha_1+\im\alpha_0$, of the partial structure factors, $S_{ij}(k)$ for $q=1$ and $\Delta=0.1$. Only the poles with $\alpha_1\geq0$ are shown, and $\alpha_0$ and $\alpha_1$ are scaled by the diameter of the particles, $\sigma_{11}=\sigma_{22}$. The poles are labelled with an (arbitrary) index, $n=0$ to 6. The parts labelled (a) to (d) correspond to the points marked in the phase diagram, figure~\ref{fig:pd_R1.0_D0.1}, (a) $\eta_1=\eta_2=0.08$, (b) $\eta_1=\eta_2=0.1$, (c) $\eta_1=\eta_2=0.12$, and (d) $\eta_1=\eta_2=0.14$. As $\eta=\eta_1+\eta_2$ increases, the imaginary component, $\alpha_0$, of the purely imaginary, $n=0$, pole decreases and the leading order pole (the one with the smallest $\alpha_0$) changes from the pair of complex, $n=1$, poles ($\Box$) to the single imaginary, $n=0$, pole ($\fullcircle\!\!$). The crossover occurs at the statepoint $\eta_1=\eta_2=0.117$.}
 \end{figure}

To elucidate the intermediate and asymptotic, $r\to\infty$, decay of $g_{ij}(r)$, the inset of figure~\ref{fig:g_r_q1} shows $\ln|rh_{ij}(r)|$. Recall that $h_{ij}(r)=g_{ij}(r)-1$ is the total correlation function.
%For statepoints close to the axes in figure~\ref{fig:pd_q1.0_d-.-} we find that the decay is damped oscillatory with a wavelength $\sim\sigma_{11}$.
For the low density case, ($i$), the intermediate and asymptotic decay is oscillatory with a wavelength $\sim\sigma_{11}$. As the statepoint approaches the coexistence region, the decay of $g_{ij}(r)$ starts to become monotonic. For ($ii$) the intermediate decay is oscillatory, but the amplitude of these oscillations quickly decreases with increasing distance $r$ and the decay becomes monotonic. For ($iii$) the oscillatory contribution decays much more rapidly and the asymptotic and even intermediate ($r\gtrsim7\sigma_{11}$) decay is monotonic.

\begin{figure}[t]
\centering
\includegraphics[width=8.5cm]{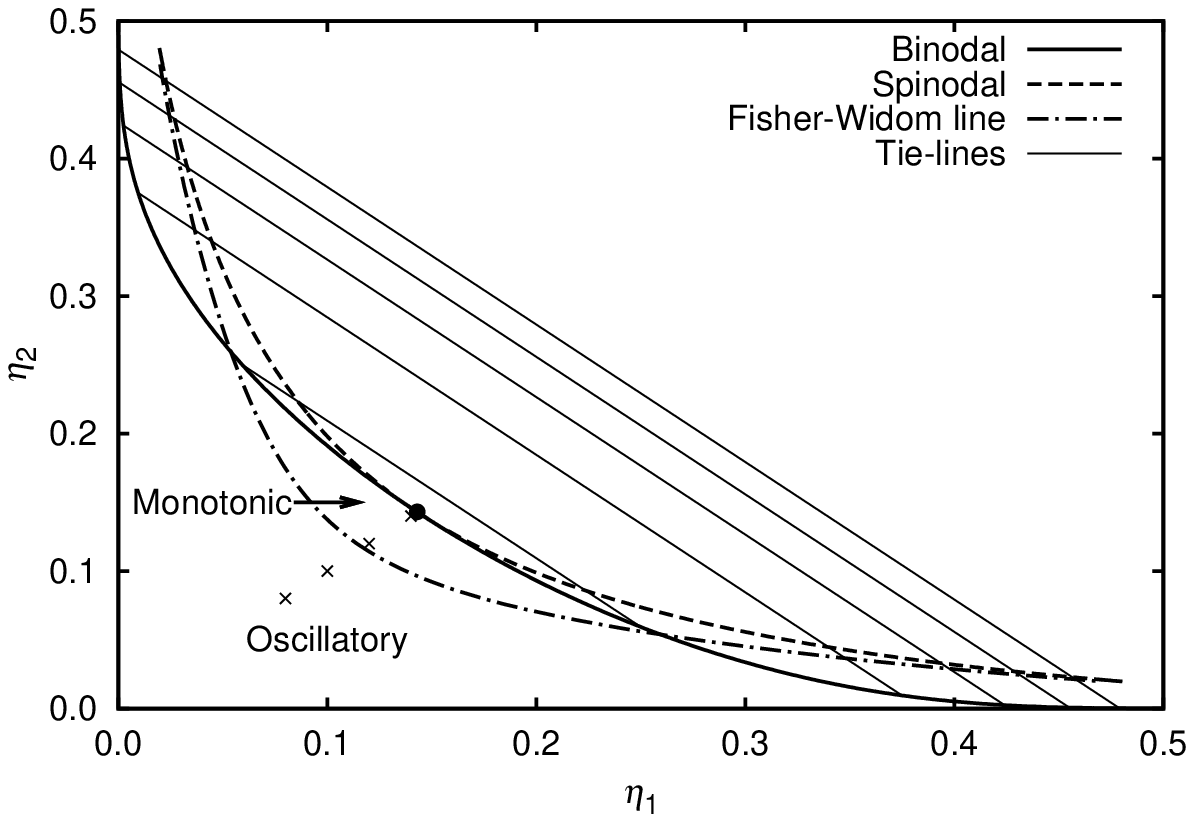}
\caption[Phase Diagram, $q=1.0$, $\Delta=0.1$.]{\label{fig:pd_R1.0_D0.1}
Fluid-fluid demixing phase diagram for $q=1$ and $\Delta=0.1$. The tie-lines are at pressures $ P\sigma_{11}^3/k_BT=3,5,7,9,11$ (from bottom to top). The Fisher-Widom line (dash-dotted line) separates regions of the phase diagram where the decay of correlation functions has a different type (monotonic or damped oscillatory) of asymptotic, $r\to\infty$, decay.}
% Bulk fluid-fluid demixing phase diagram for the binary non-additive hard sphere fluid with size ratio $q=\sigma_{11}/\sigma_{22}=1$ and non-additivity parameter $\Delta=0.1$, plotted as a function of the partial packing fractions, $\eta_1$ and $\eta_2$. The binodal (solid line) meets the spinodal (dashed line) at the bulk critical point ($\bullet$). The dotted lines are tie-lines with pressures $ P\sigma_{11}^3/k_BT=3$ to 11 in increments of 2. The Fisher-Widom line (dash-dotted line) separates regions of the phase diagram where the decay of particle correlation functions has different modes (monotonic and damped oscillatory) of asymptotic, $r\to\infty$, decay.}
%Away from the critical point the ultimate decay is damped oscillatory with a wavelength similar to $\sigma_{11}$ but as the spinodal is approached there is an abrupt crossover to monotonic, exponentially decreasing, decay. %The `$\square$' denote the statepoints used to calculate the fluid interfaces in Fig.~\ref{fig:fi_R1.0_D0.1}
 \end{figure}

Calculating $g_{ij}(r)$ for the asymmetric case $q<1$ with $\Delta>0$ we find that there are three types of intermediate and asymptotic, $r\to\infty$, decay. In figure~\ref{fig:g_r_q2} we plot representative examples of $g_{ij}(r)$ corresponding to the three types of decay for parameters $q=0.5$ and $\Delta=0.1$. In the main panel of figure~\ref{fig:g_r_q2} we show the set of $g_{ij}(r)$ for the three statepoints ($i$) $\eta_1=0.151$, $\eta_2=0.001$, ($ii$) $\eta_1=0.051$, $\eta_2=0.051$, and ($iii$) $\eta_1=0.151$, $\eta_2=0.07$. On the linear scale there is relatively small variation in the overall magnitude of the correlation functions between the three statepoints. In the subplots of figure~\ref{fig:g_r_q2} we display $\ln|rh_{ij}(r)|$ for the same statepoints. For statepoints close to the $\eta_1$-axis ($i$), we find that the decay is oscillatory with a wavelength $\sim\sigma_{11}$. For statepoints which are close to the $\eta_2$-axis ($ii$), the intermediate and ultimate decay is oscillatory with wavelength $\sim\sigma_{22}$. Since $\sigma_{22}=2\sigma_{11}$ we find that the oscillatory wavelength in ($ii$) is approximately twice as large as that in ($i$). Again, as we approach the coexistence region ($iii$), we find that the intermediate decay is oscillatory, but the relative amplitude of the oscillations decreases with increasing $r$ and the ultimate decay is monotonic.

In order to understand this behaviour, we next  determine the pole structure. For the one-component hard sphere fluid ($q=1$, $\Delta=0$), it has been established that for all statepoints there is an infinite number of complex poles, $k_n=\alpha_1+\im \alpha_0$, but there are no purely imaginary poles~\cite{grodon2005hai}. Therefore, the asymptotic decay of the distribution functions, which is determined by the pole(s) with the smallest imaginary part, $\alpha_0$, will always be damped oscillatory, $rh_{ij}(r)\sim 2A_{ij} \exp( -\alpha_0r) \cos(\alpha_1 r - \theta_{ij})$ and since $q=1$ there is only one length-scale in the fluid, so the oscillatory wavelength, $2\pi/\alpha_1$, is always $\sim\sigma_{11}$.

\begin{figure}[t]
\centering
\includegraphics[width=8.5cm,height=5.5cm]{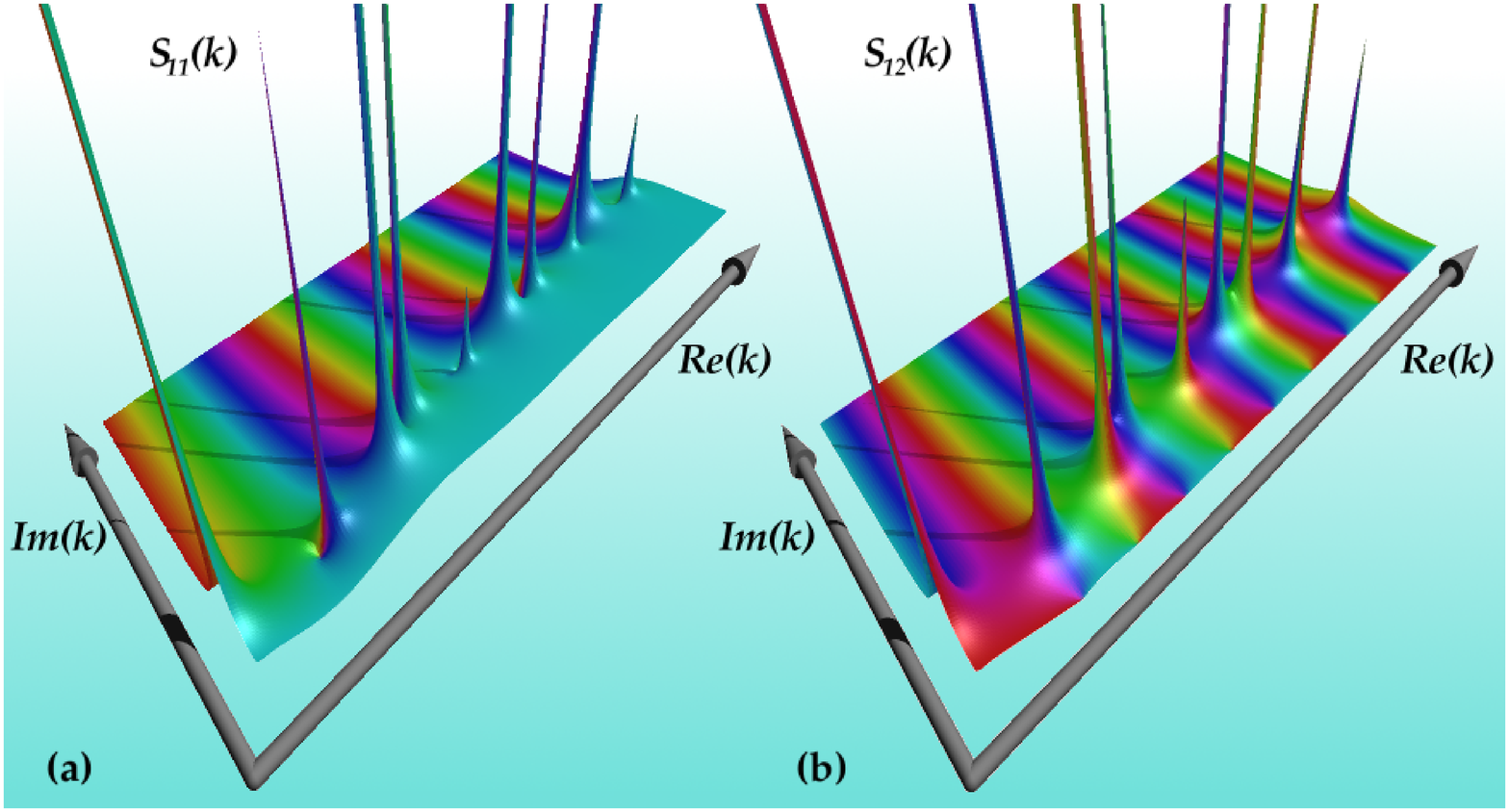}\\
\includegraphics[width=4.25cm,height=5.5cm]{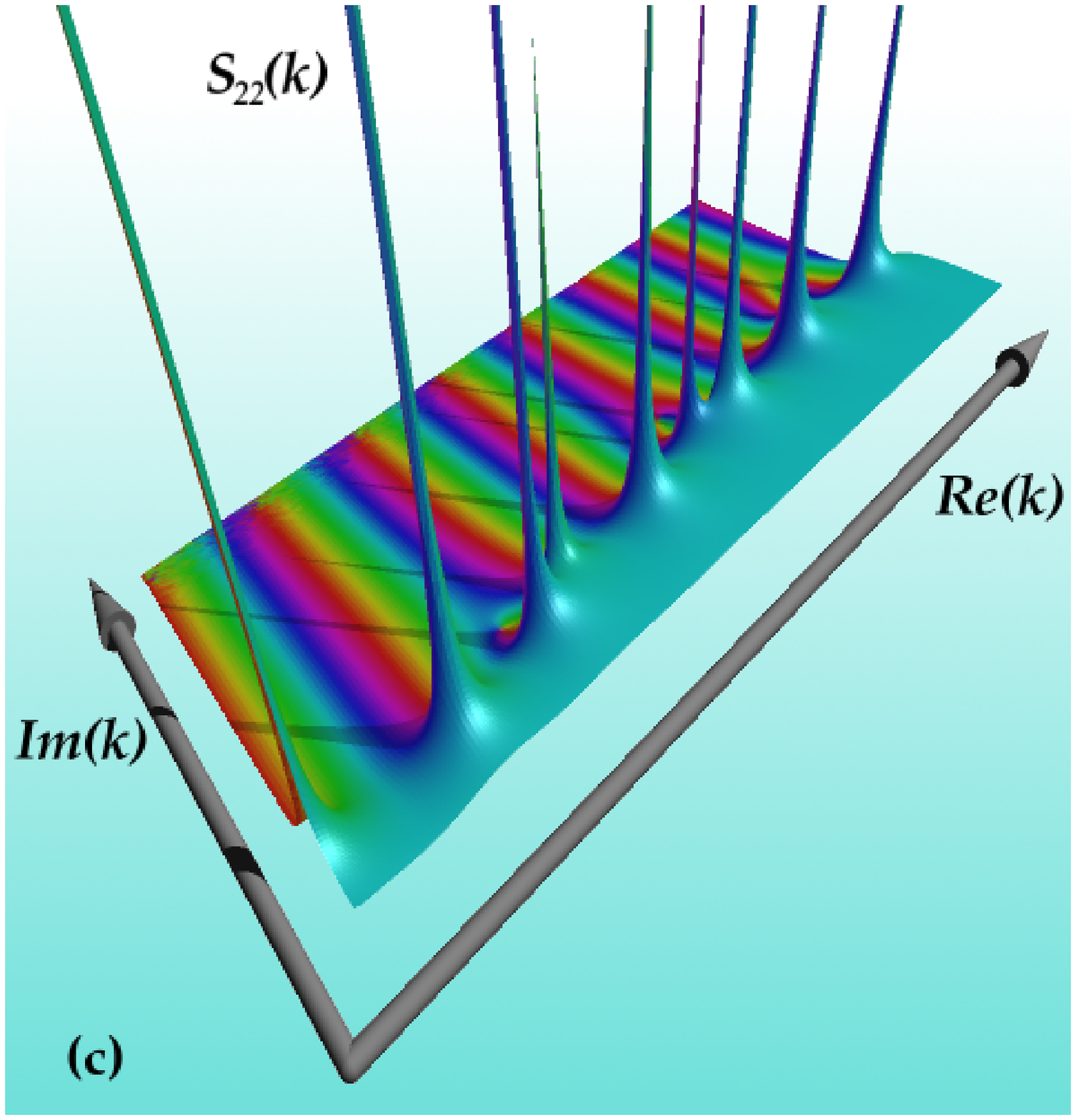}
\includegraphics[width=4.25cm]{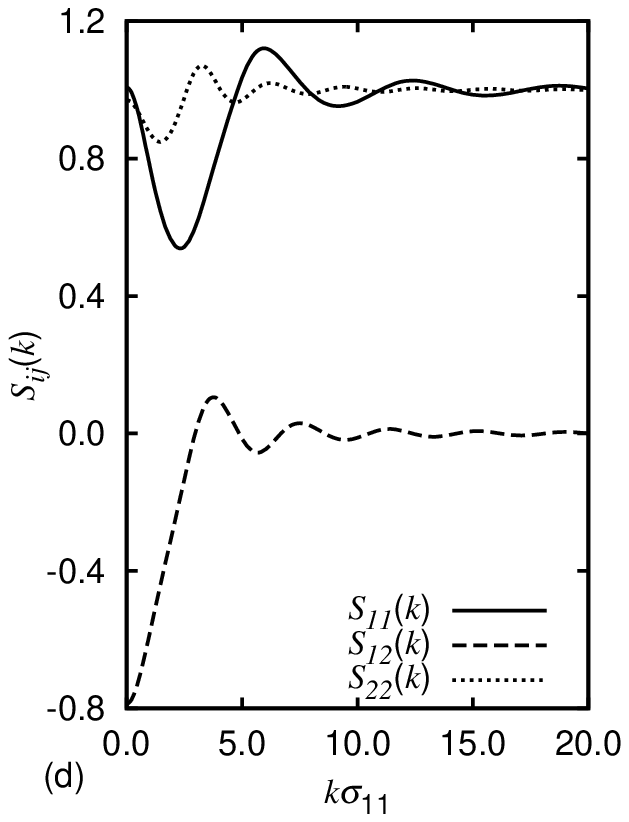}
\caption[3D Structure Factors.]{\label{fig:cmplx_strc_fctr2}
Same as figure~\ref{fig:cmplx_strc_fctr}, but for $q=0.5$ and $\Delta=0.1$, at statepoint $\eta_1=0.111$, $\eta_2=0.051$. The surfaces are again plotted over the range $k=0$ to $k=20+10\im$. Since the mixture is asymmetric, the intra-species structure factors, which are plotted in (a) and (c), are no longer identical. The real partial structure factors are shown in (d).}
 \end{figure}

For the symmetric mixture, $q=1$, but with non-additivity $\Delta>0$, we find an infinite number of complex poles, as in the additive case, but there is also a single purely imaginary pole. In order to illustrate how the poles appear in the complex structure factors, in figure~\ref{fig:cmplx_strc_fctr} we plot $S_{11}(k)=S_{22}(k)$ and $S_{12}(k)$ as a function of the complex wave-number $k$, for parameters $q=1$, $\Delta=0.1$, and statepoint $\eta_1=\eta_2=0.12$. The height and the colour of the surface plots represents the amplitude and the polar argument, respectively, of $S_{ij}(k)$. Although $S_{11}(k)$ and $S_{12}(k)$ are very different, they both exhibit sharp divergences at identical positions in the complex plane. These are the common poles of the complex partial structure factors. They are located at solutions of the equation, $\hat{D}(k)=0$, where $\hat{D}(k)$ is the common denominator~\eref{eq:denom} of the complex structure factors. To determine the positions of the poles, we numerically solve $\hat{D}(k_n)=0$ for complex $k_n=\alpha_1+\im \alpha_0$. The relationship between the complex structure factor(s) and their more commonly known real structure factor(s) is that the former evaluated along the real axis equals the latter, see figure~\ref{fig:cmplx_strc_fctr}(c).

 \begin{figure}[t]
\centering
\includegraphics[width=8.5cm]{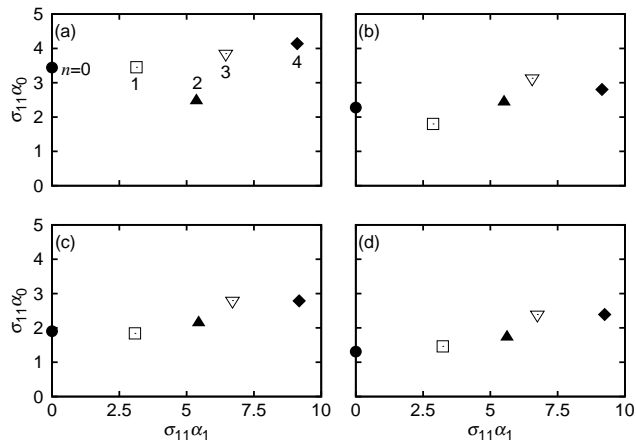}
\caption[Phase Diagram, $q=1.0$, $\Delta=0.1$.]{\label{fig:poles_R2.0}
Same as figure~\ref{fig:poles_R1.0}, but for $q=0.5$ and $\Delta=0.1$. The parts labelled (a) to (d) correspond to the points marked in the phase diagram, figure~\ref{fig:pd_q2.0_d0.1}, (a) $\eta_1=0.151$, $\eta_2=0.001$, (b) $\eta_1=0.051$, $\eta_2=0.051$, (c) $\eta_1=0.111$, $\eta_2=0.051$, and (d) $\eta_1=0.151$, $\eta_2=0.051$. In part (a) the leading order poles are a complex pair, $n=2$, ($\blacktriangle$). In (b) and (c) the leading order poles are a different complex pair, $n=1$, ($\Box$), and in part (d) the leading order pole is the purely imaginary pole, $n=0$ ($\fullcircle\!\!$).}
 \end{figure}

Figure~\ref{fig:poles_R1.0} displays a sequence of positions of the poles in the complex plane as the statepoint approaches the coexistence region, as indicated in the phase diagram in figure~\ref{fig:pd_R1.0_D0.1}. We show the positions of six of the complex poles, with index $n=1$ to 6 (arbitrarily) labelling the poles, along with the single imaginary pole, $n=0$. For each conjugate complex pair of poles, $k=\pm\alpha_1+i\alpha_0$ we show only the pole with real part $\alpha_1>0$. In part (a), at statepoint $\eta_1=\eta_2=0.08$, the complex pair, $n=1$, of poles with real part $\sigma_{11}\alpha_1\sim2\pi$ are the leading order poles and give rise to ultimate oscillatory decay with a wavelength $2\pi/\alpha_1\sim\sigma_{11}$. As we increase the total density, the positions of all poles change, but in general their imaginary components, $\alpha_0$, decrease, see figure~\ref{fig:poles_R1.0}(b) for $\eta_1=\eta_2=0.1$.

The decrease in the value of $\alpha_0$ proceeds much more rapidly for the purely imaginary pole, $n=0$, than for the complex poles and for $\eta>0.117$ this pole possesses the smallest imaginary part and therefore becomes the leading order pole. Figure~\ref{fig:poles_R1.0}(c) is at statepoint $\eta_1=\eta_2=0.12$, where the leading order pole is the ($n=0$) purely imaginary pole. This determines the ultimate asymptotic decay of correlations to be purely exponential. Increasing $\eta$ further results in the imaginary components of all poles decreasing. In part (d), at statepoint $\eta_1=\eta_2=0.14$, which is very close to the bulk critical point, the purely imaginary pole, $n=0$, is still the leading order pole, and is very close to the real axis. As the critical point (or in general the spinodal) is approached the purely imaginary pole approaches the real axis, which corresponds to the divergence of the correlation length, $1/\alpha_0$.

By varying $\eta_1$ and $\eta_2$ we determine the statepoints where the leading order pole(s) changes from the  $n=1$ complex pair of poles to the $n=0$ purely imaginary pole. This yields the FW crossover line. We find that the FW line lies between the spinodal and the axes, and that the two ends of the FW line approach the spinodal -- see figure~\ref{fig:pd_R1.0_D0.1} which displays the FW line alongside the binodal and spinodal for the case $q=1$ and $\Delta=0.1$. Note that the FW line intersects the binodal twice. Since the mixture is symmetric ($q=1$) this occurs at coexisting statepoints. We show the importance of this feature below, when we investigate the planar fluid-fluid profiles.

% If we return to the additive model, but for an asymmetric size ratio ($q<1$ and $\Delta=0$) it has been found previously that there is another set of complex poles, which are related to the new length-scale, $\sigma_{22}$~\cite{grodon2005hai}. Since all of the poles are complex the decay mode must always be damped oscillatory, but the wavelength of these oscillations, related to the real part of the leading order pole, can now be approximately one of two values; it can either be similar to the diameter of the larger species, or similar to the diameter of the smaller species. In other words, the damped oscillatory decay present in all of the partial correlation functions is either determined by the smaller or the larger particles. As the parameters and statepoint are varied the poles move about the complex plane and the pole which is currently the leading order pole can change from one determining a smaller wavelength to one determining a larger wavelength. Since the two poles must have the same imaginary part at the crossover, the asymptotic decay length of the correlation functions does not change, but the oscillatory wavelength jumps abruptly.

 \begin{figure}[t]
\centering
\includegraphics[width=8.5cm]{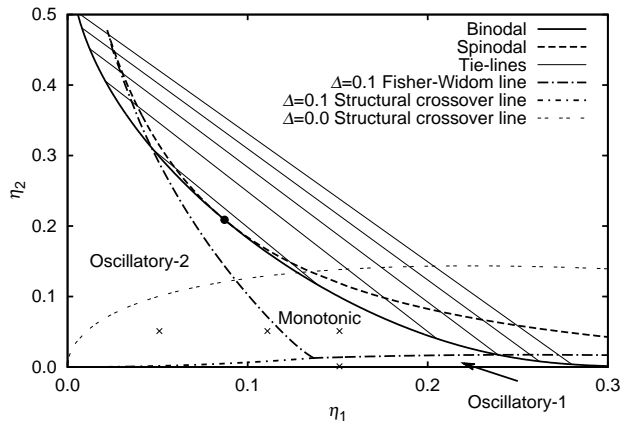}
\caption[Phase Diagram, q=2.0, $\delta=0.1$.]{\label{fig:pd_q2.0_d0.1}
Same as figure~\ref{fig:pd_R1.0_D0.1}, but for size ratio $q=0.5$ and $\Delta=0.1$. The tie-lines are shown at pressures $P\sigma_{11}^3/k_BT=1,1.25,1.5,1.75$ and $2$ (from bottom to top). There are three regions of the phase diagram, each with its own type of asymptotic decay. The structural crossover line (short-dash dotted line) separates the two regions with oscillatory decay and the Fisher-Widom line (long-dash dotted line) separates the two regions where the asymptotic decay is oscillatory from the region where the decay is monotonic. The structural crossover line for the additive case, $\Delta=0$ (double dashed line), is shown for comparison.}
% The oscillatory region close to the  $\eta_1$-axis (Oscillatory$_1$) has a wavelength similar to $\sigma_{11}$, and the second oscillatory region (Oscillatory$_2$) has a wavelength similar to $\sigma_{22}$. }
\end{figure}

\begin{figure}[t]
\centering
\includegraphics[width=8.5cm]{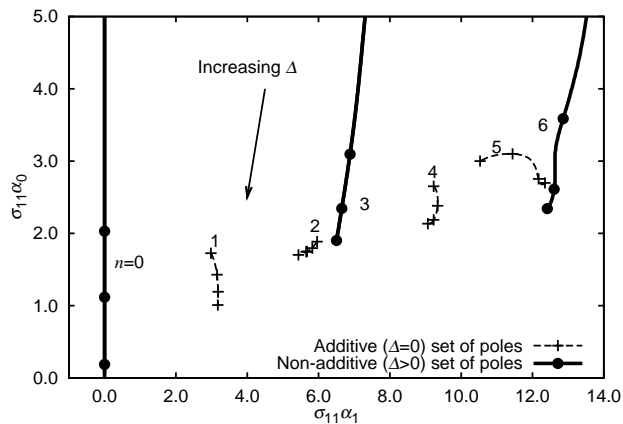}
\caption[Poles, $q=2.0$, $\eta_1=\eta_2=0.1$, $\delta=0.0$ to $0.17$.]{\label{fig:poles_R2.0_vardelta}
The positions, $k_n=\alpha_1+\im\alpha_0$, of the poles of the structure factors, $S_{ij}(k)$, in the complex plane for $q=0.5$, $\eta_1=\eta_2=0.1$ and increasing $\Delta$ from 0 to 0.164. The positions of the poles are indicated for $\Delta=$ 0, 0.054, 0.108, 0.162. At $\Delta=0$ there exists an infinite set of complex poles (no purely imaginary poles), four of which are shown here, $n=1,2,4$ and 5 ($+$). The asymptotic decay is necessarily damped oscillatory, determined by the $n=1$ pair of conjugate complex poles. As $\Delta$ is increased from zero, a second set of poles, three of which are shown $n=0,3,$ and 6 ($\newmoon$), including one purely imaginary pole ($n=0$), appears.  Initially these poles have large imaginary components i.e.~large $\alpha_0$.  As $\Delta$ is increased, the imaginary components, $\alpha_0$, of the new set of poles decreases (the poles move down the complex plane). At $\Delta=0.101$ the leading order pole changes from the $n=1$ pair of poles in the original set to the purely imaginary, $n=0$, pole in the new set, via FW crossover. The asymptotic decay is now monotonic. As $\Delta$ is increased further the value of $\alpha_0$ of the purely imaginary, $n=0$, pole then decreases to zero which is equilvalent to the correlation length diverging at the spinodal.}
%
% Therefore, the leading order pole must be complex and thus gives rise to damped oscillatory asymptotic decay of correlation functions, in this instance with a wavelength $\sim\sigma_{11}$. , resulting in the asymptotic decay of the correlation functions changing from damped oscillatory to exponential asymptotic decay. }
\end{figure}

Turning to the asymmetric non-additive mixture, we find that there is again an infinite number of complex poles with non-vanishing real parts, as well as a single purely imaginary pole, see figure~\ref{fig:cmplx_strc_fctr2}.
% For statepoints close to the axes the leading order poles are one of the pairs of complex poles. This gives rises to ultimate oscillatory decay with a wavelength similar to the diameter of the majority component.
In figure~\ref{fig:poles_R2.0} we plot the positions of the poles, with (arbitrary) index $n=0$ to 4, at four different statepoints for parameters $q=0.5$, and $\Delta=0.1$, as indicated in the phase diagram in figure~\ref{fig:pd_q2.0_d0.1}. Figure~\ref{fig:poles_R2.0}(a) plots the positions of the poles for statepoint $\eta_1=0.151, \eta_2=0.001$, which is very close to the $\eta_1$-axis. The leading order poles are a complex pair, $n=2$, that give rise to ultimate oscillatory decay with a wavelength, $\lambda=2\pi/\alpha_1\sim1.2\sigma_{11}$. Figure~\ref{fig:poles_R2.0}(b) is at statepoint $\eta_1=0.051, \eta_2=0.051$ where
% and although this statepoint has a significant value of $\eta_2$
the leading order poles are a different complex pair, $n=1$, that gives rise to an oscillatory wavelength, $\lambda=2.83\sigma_{11}$. As the coexistence region is approached, the imaginary components of all the poles decreases, see figure(c), which is at statepoint $\eta_1=0.111, \eta_2=0.051$. This decrease in the imaginary components proceeds most rapidly for the purely imaginary, $n=0$, pole which in figure~\ref{fig:poles_R2.0}(d) ($\eta_1=0.151, \eta_2=0.051$), now possesses the smallest value of $\alpha_0$ and thus becomes the leading order pole.

Therefore, for $q<1$ and $\Delta>0$ by varying the statepoint we find {\it three} regions of the phase diagram that have different types of asymptotic decay. There are two regions, close to the axes, where the decay is damped oscillatory with a wavelength similar to the majority component, and there is one region, which contains the spinodal and critical point, with monotonic decay. The phase diagram in figure~\ref{fig:pd_q2.0_d0.1} shows these regions for the mixture with $q=0.5$ and $\Delta=0.1$. Note that the FW line again crosses the binodal twice, but that these crossings do not occur in coexisting phases. Thus, there are coexisting phases which have different types of asymptotic decay, or different oscillatory decay wavelengths.

The most obvious difference that distinguishes the non-additive from the additive mixture is the presence of an additional purely imaginary pole in the former case. In order to understand where this pole comes from, and to elucidate the full effect of introducing non-additivity on the pole structure, we start with an additive mixture and slowly increase $\Delta$ from zero. Figure~\ref{fig:poles_R2.0_vardelta} displays the positions of the poles for fixed $q=0.5$, $\eta_1=\eta_2=0.1$ and increasing $\Delta$ from zero. The asymmetric additive hard sphere mixture has two sets of complex poles, each set having real components that are related to each of the length-scales. As $\Delta$ is increased from zero, a third set of complex poles, including the single purely imaginary pole, with very large imaginary components appear. As $\Delta$ increases, the imaginary components, $\alpha_0$, of the new set of poles decrease so that this set of poles moves into positions in the complex plane comparable to the set of original `additive' poles. Increasing $\Delta$ further results in the decrease of the imaginary parts of the new set of `non-additive' poles. This proceeds most rapidly for the purely imaginary pole, $n=0$, (from the new set of `non-additive' poles) which then becomes the leading order pole. For these parameters, the FW crossover occurs at $\Delta=0.101$. As $\Delta$ is increased further, the purely imaginary pole reaches the real axis, which corresponds to the divergence of the correlation wavelength at the spinodal. In the following section, we will investigate the repercussion of the asymptotic decay of correlation on the structure of the free fluid interface.

\subsection{Structure of the Free Fluid-Fluid Interface}
\label{sec:res_fi}

\begin{figure}[t]
\centering
\includegraphics[width=8.5cm]{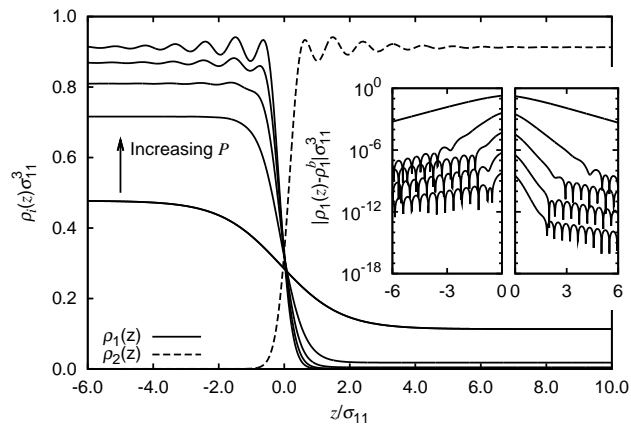}
\caption[Phase Diagram, q=2.0, $\delta=0.1$.]{\label{fig:fi_R1.0_D0.1}
Free interface density profiles, $\rho_i(z)$, between coexisting fluid phases for $q=1$ and $\Delta=0.1$, and pressures $P\sigma_{11}^3/k_BT=3$, 5, 7, 9 and 11 corresponding to the tie-lines in figure~\ref{fig:pd_R1.0_D0.1}. The profiles are plotted as a function of scaled position from the interface, $z/\sigma_{11}$. Since the mixture is symmetric, the density profiles of species 2 are identical to those of species 1 under the reflection $z\to-z$. The insets show the decay of $|\rho_1(z)-\rho_1^{b}|$ for $z<0$ and $z>0$ where $\rho_1^b$ is the bulk density of species 1 on the side of the interface shown in each inset. The profiles in the insets are plotted on a semi-logarithmic scale and each profile is offset from the one above by a factor of $10^{-2}$.}
% At low pressures the profiles of appear monotonic but as the pressure increases they exhibit proncouned oscillations.  The pair of coexisting statepoints for the profile at the lowest pressure reside within the region of the phase diagram with monotonic asymptotic decay, therefore the asymptotic decay of the profiles is thus monotonic. The pairs of coexisting statepoints for the remaining profiles are all within the region of the phase diagram with damped oscillatory asymptotic decay therefore the asymptotic decay of the profiles must be oscillatory.}
\end{figure}

\begin{figure}[t]
\centering
\includegraphics[width=8.5cm]{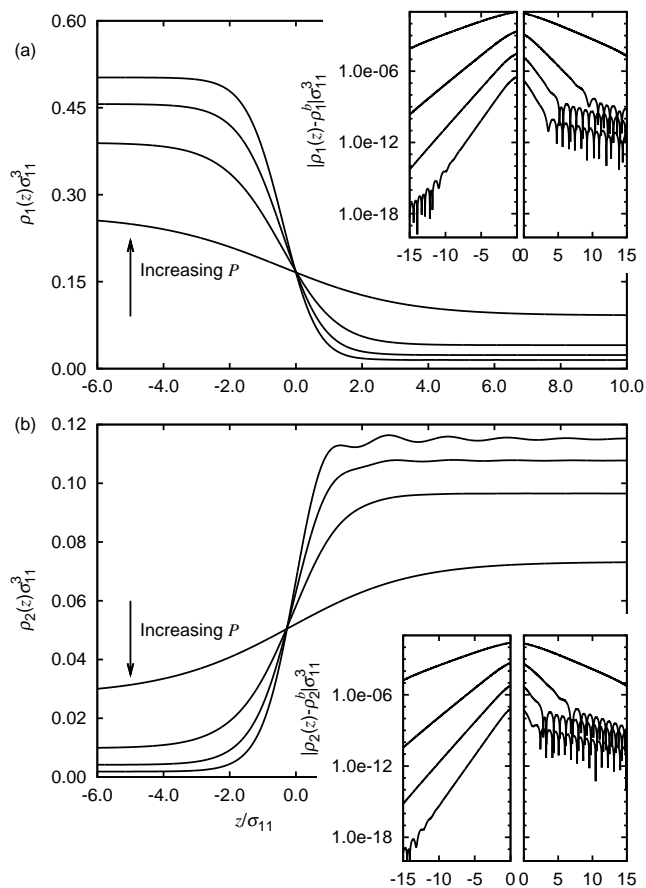}
\caption[Phase Diagram, q=2.0, $\delta=0.1$.]{\label{fig:fi_R2.0_D0.1}
Same as figure~\ref{fig:fi_R1.0_D0.1}, but for $q=0.5$ and $\Delta=0.1$. The profiles for the smaller particles are plotted in (a) and those for the larger, species 2, in (b). The coexisting pressures are $P\sigma_{11}^3/k_BT=1.0$, 1.25, 1.5, and 1.75 corresponding to the tie-lines in figure~\ref{fig:pd_q2.0_d0.1}. The insets show the asymptotic decay of the profiles on a semi-logarithmic scale, where each profile is offset by a factor $10^{-2}$ from the one above.
% On a normal scale most of the profiles appear to be monotonic, the only exception being for very highest pressures where $\rho_2(z)$ starts to appear oscillatory for $z>0$.  At the lowest pressure the profiles for both species decay monotonically on both sides of the interface. The profiles for asympototic decay becomes oscilatory 
%which has a state-point rich in species two in the oscillatory region and the statepoint poor in species 1 in the monotonic region. (b) $\beta P\sigma_{11}^3=1.72$ which has both statepoints in the oscillatory region. The inset plots the absolute difference between the profiles and the bulk densities, $\rho_i\suprm{b}=\rho_i(z\to\pm\infty)$, on that side of the interface. Note that the insets are plotted on a log scale. Although it is not obvious to discern how the density profiles decay by inspecting the main figures we see that the ultimate decay is as predicted from the particle correlation functions.
}
\end{figure}

By numerically minimising $\Omega[\rho_1,\rho_2]$, using the method outlined in section~\ref{sec:inhomo_systems} and the planar weight functions given in~\ref{sec:weifncs_1d}, we calculate the equilibrium one-body density profiles, $\bar{\rho}_i(\rr)$ (referred to as $\rho_i(\rr)$ in the following) for the fluid-fluid interface between coexisting phases with a simple iterative Picard scheme.
%In order to demonstrate these three regimes we choose the particular parameters, $q=1.0$, and  $q=2$, both with $\Delta=0.1$,  and ~\ref{fig:fig:pd_q2.0_d0.1}. 
%For $q=1$ and $\Delta=0.1$, we showed that there are only two asymptotic decay modes and that the Fisher-Widom line separating the regions of the phase diagram with these two decay modes must therefore intersect the coexistence curve on both sides of the demixed region at the same pressure.
Figure~\ref{fig:fi_R1.0_D0.1} shows the  density profiles, $\rho_i(z)$, of the free interface as a function of distance $z$ from the interface for a range of pressures, corresponding to the tie-lines shown in figure~\ref{fig:pd_R1.0_D0.1}. Since the mixture is symmetric, we plot all results for species 1, but only one representative profile for species 2 as an illustration. Starting with coexisting phases close to the critical point, we find that the density profiles vary monotonically as a function of $z$. This is fully consistent with the type of asymptotic decay of the pair correlation functions, which is monotonic for both coexisting statepoints. As one moves away from the critical point one finds that $\rho_1(z)$ becomes oscillatory on the side of the interface where species 1 is the majority component ($z<0$). This agrees with our results for the asymptotic decay changing on crossing the FW line, but there is apparently no oscillations on the side of the interface where species 1 is the minority component ($z>0$). Similarly, $\rho_2(z)$ is oscillatory at $P\sigma_{11}^3/k_BT=11$ for $z>0$ but does not appear to be oscillatory for $z<0$. 

It is clear that as the pressure is increased, the oscillations in the density profiles for each species appear on one side of the interface, but to examine how the abrupt change in the type of asymptotic decay on crossing the FW line affects the profiles, we must investigate the decay of the profiles away from the interface. The inset in figure~\ref{fig:fi_R1.0_D0.1} shows the intermediate decay of $\rho_1(z)$ for $z<0$ and for $z>0$. We plot the absolute difference of the density profiles and their bulk value on that side of the interface, $|\rho_1(z)-\rho_1^b|$, on a logarithmic scale. The profile between the coexisting phases closest to the critical point, $P \sigma_{11}^{3}/(k_BT)=3$, clearly decays monotonically on both sides of the interface. The case $P \sigma_{11}^{3}/(k_BT)=5$, appears to be monotonic on the linear plot, but if one looks at the intermediate decay behaviour (shown in the inset), one finds that there is oscillatory decay on both sides of the interface, but that these oscillations do not appear until at least a distance $z/\sigma_{11}\simeq3$ from the interface. As the pressure of the coexisting phases is increased, the oscillations shown in the insets grow in relative amplitude and start to appear closer to the interface and thus become more pronounced on the linear scale.

%  so that the amplitude of oscillations is of a similar order to the difference in bulk densities. Fig.~\ref{fig:fi_R2.0_D0.1} plots the profiles for a very high fluid density corresponding to the statepoints ($\eta\suprm{A}_1=0.49,\eta\suprm{A}_2=1.2\times10^{-4}$) and ($\eta\suprm{B}_1=1.2\times10^{-4},\eta\suprm{B}_2=0.49$).

% For $q<1$ we showed that there are up to three modes of asymptotic behaviour, one monotonic region and two regions with damped oscillatory decay with different wavelengths.
For $q=0.5$ and $\Delta=0.1$, corresponding to the phase diagram shown in figure~\ref{fig:pd_q2.0_d0.1}, we again start at the critical point and trace pairs of coexisting state-points along the binodal. Figure~\ref{fig:fi_R2.0_D0.1} displays the density profiles for coexisting phases, corresponding to the tie-lines in figure~\ref{fig:pd_q2.0_d0.1}. Close to the critical point the coexisting state-points both reside in the region of the phase diagram where the asymptotic decay of $g_{ij}(r)$ is monotonic. On the linear plot we find that the density profiles for coexisting phases close to the critical point appear monotonic. The inset of figure~\ref{fig:fi_R2.0_D0.1}(a) shows that the density profile corresponding to the lowest pressure, $P\sigma_{11}^3/k_BT=1$, decays monotonically on both sides of the interface. As we increase the pressure and move along the coexistence curve, we find that the state-points rich in species 2 crosses the FW line and moves into the oscillatory region (labelled Oscillatory-2), while the other state-point remains in the monotonic region. Therefore, the density profiles decay with an oscillatory component on the side of the interface where species 2 is the majority component ($z>0$). The inset shows that the profiles for $P\sigma_{11}^3/k_BT=1.25$ and 1.5 both exhibit this behaviour; for $z<0$ the decay is monotonic and for $z>0$ the decay is oscillatory. If the pressure is increased further, the other state-point (rich in species 1) crosses the FW line and moves into the other oscillatory region (Oscillatory-1). The inset shows that the intermediate decay of the profiles for $P\sigma_{11}^3/k_BT=1.75$ is oscillatory, but that this is very far away from the interface and occurs with a small relative amplitude. These coexisting phases both have oscillatory decay but with different wavelengths; if one examines the profile for $P\sigma^3/k_BT=1.75$, the oscillatory wavelength for $z>0$ is approximately twice as large as that for $z<0$.

\subsection{Interface Tension of the fluid-fluid interface}

\begin{figure}[t]
\centering
\includegraphics[width=8.5cm]{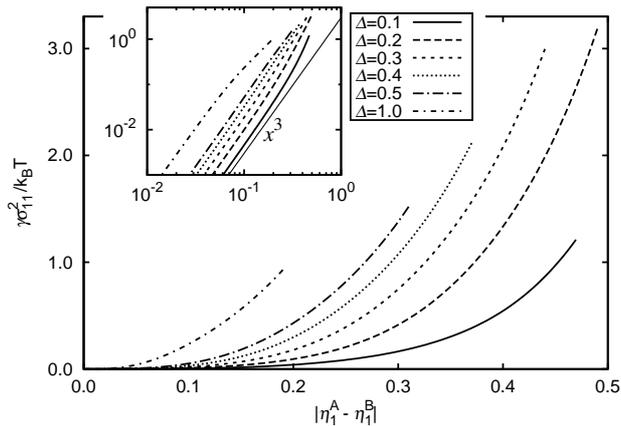}
\caption[Surface Tension, q=0.5=1.0, $\delta=0.1$.]{\label{fig:surface_tension_1.0}
The surface tension, $\gamma$, of the planar fluid-fluid interfaces with parameters $q=1$ and $\Delta=0.1$, 0.2, 0.3, 0.4, 0.5 and 1, plotted against the absolute difference in the partial packing fraction of species 1, $|\eta_1^{\rm A}-\eta_1^{\rm B}|$, in the two coexisting phases A and B. The inset shows the same quantities on a double logarithmic scale, and compares them to the mean-field behaviour, $\gamma\propto |\eta_1^{\rm A}-\eta_1^{\rm B}|^3$, labelled $x^3$ (thin solid line).
}
\end{figure}

\begin{figure}[t]
\centering
\includegraphics[width=8.5cm]{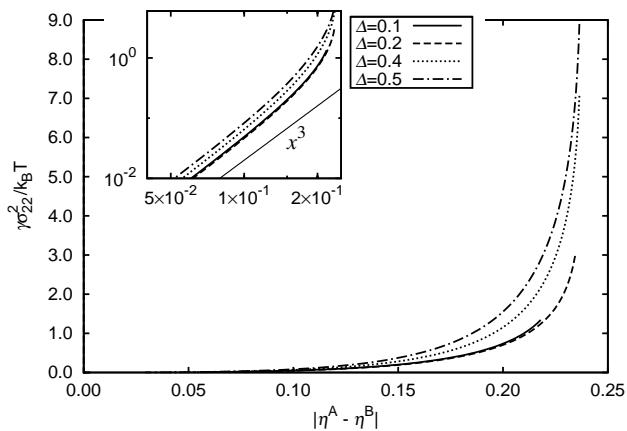}
\caption[Surface Tension, q=0.5=1.0, $\delta=0.1$.]{\label{fig:surface_tension_2.0}
Same as figure~\ref{fig:surface_tension_1.0}, but for $q=0.5$ and $\Delta=0.1$, 0.2, 0.4 and 0.5, and shown as a function of the absolute difference in the {\it total} packing fraction, $|\eta^{\rm A}-\eta^{\rm B}|$, in the two coexisting phases A and B. Within this representation we find that for fixed order parameter, $|\eta^{\rm A}-\eta^{\rm B}|$, $\gamma$ varies non-monotonically with $\Delta$. The inset shows the same quantities on a double logarithmic scale and compares them to the mean-field behaviour, $\gamma\propto |\eta^{\rm A}-\eta^{\rm B}|^3$, again labelled $x^3$.

 }
\end{figure}

From the density profiles we have calculated the surface tension of the free interface,
\begin{equation}
\gamma= (\Omega[\rho_1,\rho_2]+PV)/A,
\label{eq:gamma_def}
\end{equation}
where $\Omega[\rho_1(\rr),\rho_2(\rr)]$ is the grand potential of the inhomogeneous system with the free interface, $-PV=\Omega[\rho_1^b,\rho_2^b]$ is the grand potential of the uniform system, and $A$ is the area of the interface. Figure~\ref{fig:surface_tension_1.0} displays the surface tension for the mixture with $q=1$ and varying $\Delta$, plotted against the order parameter $|\eta_1^{A}-\eta_1^{B}|$, where $\eta_1^{A}$ is the packing fraction of species 1 in phase $A$ (and similarly for $B$). As the mixture is symmetric, this quantity is symmetric w.r.t.~interchange of species, i.e. $|\eta_1^{A}-\eta_1^{B}|=|\eta_2^{A}-\eta_2^{B}|$.
% We find that for all values of $\Delta$ the value of $\gamma$ decreases as $|\eta_1^{A}-\eta_1^{B}|$  i.e. the coexisting state-points approach the critical point.
We find that increasing the non-additivity has a dramatic effect on the surface tension, for constant $|\eta_1^{A}-\eta_1^{B}|=0.1$, $\gamma$ increases over fifty times between $\Delta=0.1$ and $\Delta=1$.

It can be shown~\cite{safran2003statistical} that as the critical point is approached, $\gamma$ follows a simple mean-field scaling law, $\gamma\propto|\eta_1^{A}-\eta_1^{B}|^3$. In order to check our calculations of $\gamma$ we plot $\gamma$ against $|\eta_1^{A}-\eta_1^{B}|$ on a double logarithmic scale in the inset of figure~\ref{fig:surface_tension_1.0}. For comparison, in the inset we show the asymptotic result, $\gamma=a|\eta_1^{A}-\eta_1^{B}|^3$ (labelled $x^3$) where $a$ is a proportionality constant. For all values of $\Delta$, as $|\eta_1^A-\eta_1^{B}|$ approaches zero, $\gamma$ tends towards the mean-field behaviour, i.e. the slope of the curves in the inset tends towards the slope of the asymptotic result.

\begin{figure}[t]
\centering
\includegraphics[width=8.5cm]{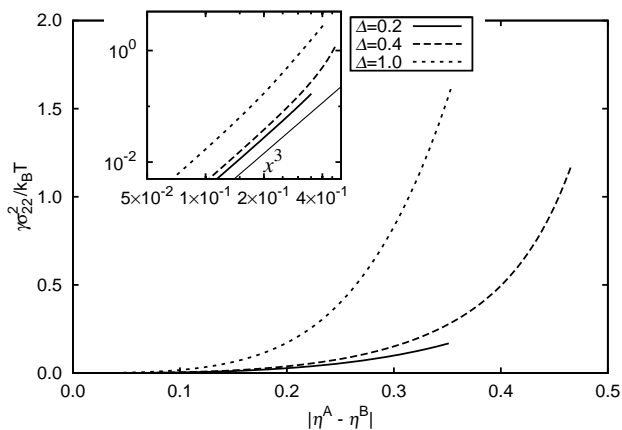}
\caption[Surface Tension, q=0.5=1.0, $\delta=0.1$.]{\label{fig:surface_tension_10.0}
Same as figure~\ref{fig:surface_tension_2.0}, but for $q=0.1$ and $\Delta=$0.2, 0.4 and 1.}
\end{figure}

Figure~\ref{fig:surface_tension_2.0} displays the surface tension of the free interface for the mixture with $q=0.5$ and varying $\Delta$. As the mixture is asymmetric, we plot these results using the (species independent) order parameter $|\eta^A-\eta^B|$ which is the absolute difference in the packing fraction between phases $A$ and $B$. Note that $\gamma$ is scaled by the square of the diameter of the larger species, $\sigma_{22}$, so as to keep the values of $\gamma$ comparable across a range of $q$ values. In this representation, we find that the curves are broadly similar to those in figure~\ref{fig:surface_tension_1.0}, but that $\gamma$ exhibits a rapid increase with increasing $|\eta^A-\eta^B|$. Furthermore, for all values of the order parameter, the value of $\gamma$ for $\Delta=0.2$ is smaller than the value for $\Delta=0.1$. These two features arise from our choice of order parameter. If we use an order parameter similar to the one in figure~\ref{fig:surface_tension_1.0}, we do not have either of these two features.  Although we have a different order parameter, the surface tension follows a similar scaling law as we approach the critical point, $\gamma\propto|\eta^A-\eta^B|^3$. In figure~\ref{fig:surface_tension_10.0} we plot $\gamma$ for the mixture with fixed $q=0.1$ and varying $\Delta$ as a function of $|\eta^A-\eta^B|$. We find that, unlike in figure~\ref{fig:surface_tension_2.0}, $\gamma$ does not increase rapidly as $|\eta^A-\eta^B|$ approaches its maximum value, and that $\gamma$ increases monotonically with $\Delta$ (for the values considered).

%In the inset to Fig.~\ref{fig:surface_tension_2.0} we plot $|\eta^A-\eta^B|$ against $\gamma$ on a double logarithmic scale, alongside the asymptotic limit, $\gamma=a|\eta_1^{A}-\eta_1^{B}|^3$, labelled, $x^3$. As the critical point is approached the slopes of the curves in the inset tend towards the asymptotic result. The rapidly increasing nature of $\gamma$ can also be seen in these plots which have a strong upturn at larger values of $|\eta^A-\eta^B|$.

%Again, the inset plots the same quantities on a double logarithmic scale and shows that the mean-field result is reacquired as $|\eta^A-\eta^B|\to0$.

\section{Discussion}
\label{sec:discus}
Using a fundamental measure density functional theory we have investigated some of the properties of homogeneous and inhomogeneous fluid states of a binary non-additive hard sphere model with positive non-additivity. This model exhibits fluid-fluid demixing. We have calculated the coexistence curves and showed that these compare reasonably well to existing simulation results. The theory predicts that the critical point occurs at a pressure and density lower than the simulation results. This is typical of mean-field type theories, such as DFT, which do not take account of all fluctuations in the fluid. We have not investigated whether the fluid-fluid phase transitions are stable with respect to crystallisation, which is expected to occur at high packing fractions. To investigate this one would require a more sophisticated functional which is capable of modelling the extreme confinement in a crystal. Moreover, even for the additive mixture, where a suitable theory exists~\cite{cuesta2002cef,roth2010fundamental}, we are not aware of any systematic DFT investigation of freezing.

Using the Ornstein-Zernike equation, we calculated the asymptotic decay of correlation functions, $g_{ij}(r)$, by solving for the poles of the partial structure factors, $S_{ij}(k)$, in the complex plane. Using Cauchy's theorem, one can express the correlation functions as an infinite sum over these complex poles. In particular the poles with the smallest imaginary part are interesting, as these determine the asymptotic, $r\to\infty$, decay of the entire set of $g_{ij}(r)$.  We find that for $q<1$ and $\Delta\geq0$ there is, at low densities, a crossover between two modes of asymptotic damped oscillatory decay with different wavelengths, which are similar to the diameters of the two species. For $\Delta>0$ we find Fisher-Widom crossover from oscillatory to monotonic asymptotic decay as the coexistence region is approached. We find that the positive non-additivity introduces a new set of poles, including one purely imaginary pole. As $\Delta$ is increased from zero this new set of complex poles appear in the complex plane initially with very large imaginary components. As $\Delta$ is increased, the value of the imaginary components decreases and the poles occupy a region of the complex plane similar to the set of poles that exist already in the additive model.

One might imagine that the new length-scale, $\sigma_{12}$, would induce a third regime where the asymptotic decay is oscillatory with a wavelength similar to the cross-species diameter, $\sigma_{12}$. However, for $\Delta>0$ this does not occur since the new set of complex poles, related to a non-zero $R_{12}$, is always accompanied by a purely imaginary pole which is always the leading order pole of this set. It would be interesting to investigate the case $\Delta<0$ in future work.
% Up unto now we have discussed the pole structure without reference to the pole amplitudes which determine the relative contribution of each pole in the density profiles and correlation functions. Although we see that there is an abrupt crossover in the decay mode this may only manifest itself in the truly asymptotic behaviour. In order to 

Furthermore, we have studied the inhomogeneous free fluid interface between coexisting phases and have calculated the density profiles and the surface tension. We showed how the type of asymptotic decay affects the intermediate and short-range behaviour of the density profiles.
%However, we do find that our numerical scheme gives rise to kinks in the profiles when the density profiles exhibit very sharp peaks for example where there is strong adsorption at a wall.
We have presented detailed results for the surface tension of the free fluid interface. These can be compared to both simulation and experimental results, and furthermore play a vital role in the investigation of capillary condensation phenomena.

\appendix
\section{Weight Functions in Fourier Space}
\label{sec:weifncs_fs}

For completeness we include the Fourier space representations of the weight functions,
\begin{eqnarray*}
\begin{array}{rclrcl}
\widetilde{w}_3 &=& 4\pi(s - k Rc)/ k^3, \quad\quad&
\widetilde{w}_2 &=& 4\pi Rs / k, \\
\widetilde{w}_1 &=&(k Rc + s )/(2k ), \quad\quad&
\widetilde{w}_0 &=& c + (k Rs /2),
\end{array}
\end{eqnarray*}
and
\begin{eqnarray*}
\begin{array}{rclrcl}
\widetilde{w}_1^\dag\;\; &=& 4\pi(k Rc + s )/ k, \quad\quad&
\widetilde{w}_0^\dag\;\; &=& c - (k Rs /2), \\
\widetilde{w}_{-1}^\dag &=& -\tfrac{1}{16\pi}(k^2 Rc + 3ks ), \quad\quad&
\widetilde{w}_{-1}   &=& (k^2 Rc - ks )/2, \\
\widetilde{w}_{-2}  &=& -\tfrac{1}{16\pi}k^3 Rs, \quad\quad&
\widetilde{w}_{-3} &=& \tfrac{1}{16\pi}(k^4 Rc - 3k^3 s ),
\end{array}
\end{eqnarray*}
where $s=\sin(kR)$ and $c=\cos(kR)$.

\section{Free Energy Density Contributions}
\label{sec:phi_terms_apx}
Representative cases of the free energy terms , $\Phi_{\alpha\beta}$, where, $\alpha,\beta=0$ to 3, are shown below. The remaining terms can be obtained through symmetry by changing the species labels: $\Phi_{\beta\alpha}=\Phi_{\alpha\beta}(n_\nu^{(1)}\to n_\nu^{(2)},n_\tau^{(2)}\to n_\tau^{(1)})$. $\eta$ is the the total packing fraction, given by $\eta=n_3^{(1)}+n_3^{(2)}$,
% \begin{widetext}
 \begin{eqnarray*}
\fl
 \Phi_{00}&=& {\frac {n_1^{ \left( 1 \right) }n_2^{
  \left( 1 \right) } \left( n_2^{ \left( 2 \right) }
  \right) ^{3}}{4 \pi  \left( 1-\eta \right) ^{4}}}+{\frac {
  \left( n_2^{ \left( 1 \right) } \right) ^{3}n_1^{
  \left( 2 \right) }n_2^{ \left( 2 \right) }}{4\pi 
  \left( 1-\eta \right) ^{4}}}+{\frac {n_0^{ \left( 1
  \right) }n_1^{ \left( 2 \right) }n_2^{ \left( 2
  \right) }}{ \left( 1-\eta \right) ^{2}}}+{\frac {n_1^{
  \left( 1 \right) }n_2^{ \left( 1 \right) }n_0^{
  \left( 2 \right) }}{ \left( 1-\eta \right) ^{2}}} +{\frac {
  n_0^{ \left( 1 \right) } \left( n_2^{ \left( 2
  \right) } \right) ^{3}}{12\pi  \left( 1-\eta \right) ^{3}}}
  \\  \fl &
  &  \quad   +2{
 \frac {n_1^{ \left( 1 \right) }n_2^{ \left( 1
  \right) }n_1^{ \left( 2 \right) }n_2^{ \left( 2
  \right) }}{ \left( 1-\eta \right) ^{3}}} +
   {\frac { \left( 
  n_2^{ \left( 1 \right) } \right ) ^{3}n_0^{ \left( 2
  \right) }}{12\pi  \left( 1-\eta \right) ^{3}}}+{\frac {
  \left( n_2^{ \left( 1 \right) } \right) ^{3} \left(  
 n_2^{ \left( 2 \right) } \right) ^{3}}{24{\pi }^{2} \left( 1-\eta
  \right) ^{5}}}+{\frac {n_0^{ \left( 1 \right) }{{ n_0}
 }^{ \left( 2 \right) }}{1-\eta}}, \\
\fl
 \Phi_{01}&=&{\frac {n_0^{ \left( 1 \right) }n_1^{ \left( 2
  \right) }}{1-\eta}}+{\frac {n_0^{ \left( 1 \right) }
  \left( n_2^{ \left( 2 \right) } \right) ^{2}}{8\pi 
  \left( 1-\eta \right) ^{2}}}+{\frac {n_1^{ \left( 1
  \right) }n_2^{ \left( 1 \right) }n_1^{ \left( 2
  \right) }}{ \left( 1-\eta \right) ^{2}}}+{\frac {n_1^{
  \left( 1 \right) }n_2^{ \left( 1 \right) } \left( {{ 
 n_2}}^{ \left( 2 \right) } \right) ^{2}}{4\pi  \left( 1-\eta
  \right) ^{3}}}+{\frac { \left( n_2^{ \left( 1
  \right) } \right) ^{3}n_1^{ \left( 2 \right) }}{12\pi 
  \left( 1-\eta \right) ^{3}}} 
\\  \fl &
  &  \quad +
{\frac { \left( n_2^{
  \left( 1 \right) } \right) ^{3} \left( n_2^{ \left( 2
  \right) } \right) ^{2}}{32{\pi }^{2} \left( 1-\eta \right) ^{4}}},  \\
\fl
 \Phi_{02}&=&{\frac {n_1^{ \left( 1 \right) }n_2^{ \left( 1
  \right) }n_2^{ \left( 2 \right) }}{ \left( 1-\eta
  \right) ^{2}}}+{\frac { \left( n_2^{ \left( 1
  \right) } \right) ^{3}n_2^{ \left( 2 \right) }}{12\pi 
  \left( 1-\eta \right) ^{3}}}+{\frac {n_0^{ \left( 1
  \right) }n_2^{ \left( 2 \right) }}{1-\eta}}, \\
\fl
 \Phi_{03}&=&{\frac { \left( n_2^{ \left( 1 \right) } \right) ^{3}
 }{24\pi  \left( 1-\eta \right) ^{2}}}+{\frac {n_1^{ \left( 1
  \right) }n_2^{ \left( 1 \right) }}{1-\eta}}-n_0^
 { \left( 1 \right) }\ln  \left( 1-\eta \right), \\
\fl
 \Phi_{11}&=&{\frac {n_1^{ \left( 1 \right) }n_1^{ \left( 2
  \right) }}{1-\eta}}+{\frac {n_1^{ \left( 1 \right) }
  \left( n_2^{ \left( 2 \right) } \right) ^{2}}{8\pi 
  \left( 1-\eta \right) ^{2}}}+{\frac { \left( n_2^{
  \left( 1 \right) } \right) ^{2}n_1^{ \left( 2 \right) }}
 {8\pi  \left( 1-\eta \right) ^{2}}}+{\frac { \left(  n_2^{ \left( 1 \right) } \right) ^{2} \left( n_2^{ \left( 2
  \right) } \right) ^{2}}{32{\pi }^{2} \left( 1-\eta \right) ^{3}}}, \\
\fl
 \Phi_{12}&=&{\frac {n_1^{ \left( 1 \right) }n_2^{ \left( 2
  \right) }}{1-\eta}}+{\frac { \left( n_2^{ \left( 1
  \right) } \right) ^{2}n_2^{ \left( 2 \right) }}{8\pi 
  \left( 1-\eta \right) ^{2}}}, \\
\fl
 \Phi_{13}&=&{\frac { \left( n_2^{ \left( 1 \right) } \right) ^{2}}
 {8\pi  \left( 1-\eta \right) }}-n_1^{ \left( 1 \right) }
 \ln  \left( 1-\eta \right), \\
\fl
 \Phi_{22}&=&{\frac {n_2^{ \left( 1 \right) }n_2^{ \left( 2
  \right) }}{1-\eta}}, \\
\fl
 \Phi_{23}&=&-n_2^{ \left( 1 \right) }\ln  \left( 1-\eta \right),  \\
\fl
 \Phi_{33}&=&  ( 1-\eta ) \ln  ( 1-\eta ) +\eta.
 \end{eqnarray*}
% \end{widetext}

% \section{Numerical Method}
% \label{sec:method}
% We briefly state how the inhomogeneous density profiles are calculated in sections~\ref{sec:res_fi} and ~\ref{sec:res_ps}. The density profiles are discretised on a one-dimensional grid with spacing $.01\sigma_{11}$. To calculate the density profiles the computational effort is focused on calculating the direct correlation functions, $c^{(1)}_i(z)$, using Eq.~\ref{eq:c1_term}. It is first necessary to calculate the set of $2\times4$ weighted densities, $n_\alpha^{(i)}(z)$, via convolutions with the weight functions. These weighted densities and then combined into the functions $\frac{\partial \Phi_{\alpha\beta}}{\partial n_\gamma^{(i)}}$, which can be reduced into a set of $2\times30$ independent functions. The first nested integral of Eq.~\eqref{c1_term} is calculated by convoluting these $2\times30$ functions with the with the set of 10 kernel weight functions, before the second nested integral is calculated by convoluting with the $2\times4$ weighted densities again. The $c^{(1)}_i(\rr)$ are then used in standard Picard iteration in order to solve for $\rho_i(z)$.

\section{Weight Functions in Planar Geometry}
\label{sec:weifncs_1d}
In this paper we consider planar density profiles, where we can simplify the convolutions by performing the integration over the radial direction in advance, yielding a set of planar weight functions;
\begin{eqnarray}
\overline{w}^{(\dag)}_\tau(z)=2\pi\int_0^\infty\dd \xi \, \xi \, w^{(\dag)}_\tau\left(\sqrt{\xi^2+z^2}\right),
\end{eqnarray}
where $\xi=\sqrt{x^2+y^2}$. This yields
%\begin{eqnarray}
%n_3(z)&=\pi\int_{z-R}^{z+R}\dd z'\rho_i(z')[R^2-(z-z')^2] \\
%n_2(z)&=2\pi R \int_{z-R}^{z+R} \dd z'\rho_i(z')
%\end{eqnarray}
%...
%Thus we get the planar weight functions,
\begin{eqnarray*}
\overline{w}_3(z)&=\pi\sgn(R)\Theta(R-|z|)(|R|^2-|z|^2),\\
\overline{w}_2(z)&=2\pi R\Theta(|R|-|z|), \\
\overline{w}_1(z)&=\tfrac{1}{4}\sgn(R)[\Theta(|R|-|z|)+z\delta(|R|-|z|)],\\
\overline{w}_0(z)&=\tfrac{3}{4}\delta(|R|-|z|)-\tfrac{1}{4}z\delta'(|R|-|z|), 
\end{eqnarray*}
and
\begin{eqnarray*}
\overline{w}_1^\dag(z)&=2\pi\sgn(R)[\Theta(|R|-|z|)+z\delta(|R|-|z|)], \\
\overline{w}_0^\dag(z)&=\tfrac{1}{4}[\delta(|R|-|z|)+z\delta'(|R|-|z|)], \\
\overline{w}_{-1}^\dag(z)&=\tfrac{1}{32\pi}[\delta'(|R|-|z|)+z\delta^{(2)}(|R|-|z|)], \\
\overline{w}_{-1}(z)&=\tfrac{3}{4}\sgn(R)\delta(|R|-|z|)+z\delta^{(2)}(|R|-|z|)], \\
\overline{w}_{-2}(z)&=\tfrac{1}{32\pi}[3\delta^{(2)}(|R|-|z|)-z\delta^{(3)}(|R|-|z|)], \\
\overline{w}_{-3}(z)&=\tfrac{1}{32\pi}\sgn(R)[-7\delta^{(3)}(|R|-|z|)+z\delta^{(4)}(|R|-|z|)].
\end{eqnarray*}
Furthermore, we can perform the convolution of a general one-dimensional function, $f(z)$, with these planar weight functions,
\begin{eqnarray*}
F^{(\dag)}_{\tau}(z)=\int\dd z'\, f(z')\overline{w}^{(\dag)}_{\tau}(|z-z'|),
\end{eqnarray*}
giving
\begin{eqnarray*}
F_3(z)&=\pi\sgn(R)\int_{z'-|R|}^{z'+|R|}\dd z' f(z')[R^2-(z-z')^2], \\
F_2(z)&=2\pi R\int_{z'-|R|}^{z'+|R|}\dd z' f(z'), \\
F_1(z)&=\tfrac{1}{4}\sgn(R)\left[\int_{z'-|R|}^{z'+|R|}\dd z' f(z')+\tfrac{R}{4}\sum_{\pm}f(z\pm |R|)\right], \\
F_0(z)&=\tfrac{1}{2}\sum_{\pm}f(z\pm|R|)-\tfrac{ R}{4}\sum_{\pm}\pm f'(z\pm |R|), 
\end{eqnarray*}
\begin{eqnarray*}
F_1^\dag(z)&=2\pi\sgn(R)\left[\int_{z'-|R|}^{z'+|R|}\dd z' f(z')+R \sum_{\pm}f(z\pm|R|)\right], \\
F_0^\dag(z)&=\tfrac{1}{2}\sum_{\pm}f(z\pm|R|)+\tfrac{R}{4}\sum_{\pm}\pm f'(z\pm|R|), \\
F_{-1}^\dag(z)&= \tfrac{3}{32\pi}\sum_{\pm}\pm f'(z\pm|R|)+\tfrac{R}{32\pi}\sum_{\pm}f''(z\pm|R|), \\
F_{-1}(z)&=\tfrac{1}{4}\sgn(R)\left[\sum_{\pm}\pm f'(z\pm|R|)-\tfrac{R}{4}\sum_{\pm}f''(z\pm|R|)\right], \\
F_{-2}(z)&=\tfrac{R}{32\pi}\sum_{\pm}\pm f^{(3)}(z\pm|R|), \\
F_{-3}(z)&=\tfrac{3}{32\pi}\sgn(R)\left[\sum_{\pm}\pm f^{(3)}(z\pm|R|)\right. \\
& \quad \quad \quad \quad \quad \quad \quad \quad \quad \quad \quad \quad+\tfrac{R}{32\pi}\sum_{\pm}f^{(4)}(z\pm|R|)\Bigg],
\end{eqnarray*}
where $f'(z)$, $f''(z)$, $f^{(3)}(z)$ and $f^{(4)}(z)$ represent successive derivatives of $f(z)$.

The weighted densities and convolutions with weight functions for the individual species are calculated using Fourier transforms, while the convolutions with the kernel functions are calculated directly by integrating over $z'$. The $\gamma$th derivative of the free energy density derivative term, $\dd \Phi_{\alpha \beta}/\dd n_\gamma$, is calculated using a central difference approximation with a symmetric ($\gamma+1$)-point stencil.

\ack
We thank R. Evans and A. Archer for useful discussions and I. Schwarz for a critical reading of the paper. We gratefully acknowledge the EPSRC for funding under grant EP/E065619/1 and the DFG for support via SFB840/A3.

\vspace{1cm}

\bibliographystyle{unsrt}
\bibliography{refs}

\end{document}